\documentclass[twocolumn]{aastex62}
\usepackage[version=3]{mhchem}
\usepackage{astro}
\usepackage{outlines}
 \usepackage{cancel}

\def \tex{\ensuremath{T_{{\rm ex}}}}
\def \ntot{\ensuremath{N_{\rm {tot}}}}

\graphicspath{{./}{figures/}}

\received{January 10, 2019}
\revised{March 25, 2019}
\accepted{\today}
\submitjournal{ApJ}

\shorttitle{Sulfur chemistry in protoplanetary disks: CS and \ce{H2CS}}
\shortauthors{Le Gal et al.}

\begin{document}

\title{Sulfur chemistry in protoplanetary disks: CS and \ce{H2CS}}

\correspondingauthor{Romane Le Gal}
\email{romane.le$\_$gal@cfa.harvard.edu}

\author[0000-0003-1837-3772]{Romane Le Gal}

\author[0000-0001-8798-1347]{Karin I. \"Oberg}
\affiliation{Harvard-Smithsonian Center for Astrophysics, 60 Garden St., Cambridge, MA 02138, USA}

\author[0000-0002-8932-1219]{Ryan Loomis}
\affil{NRAO, 520 Edgemont Rd, Charlottesville, VA 22903, USA}

\author{Jamila Pegues}
\affiliation{Harvard-Smithsonian Center for Astrophysics, 60 Garden St., Cambridge, MA 02138, USA}

\author[0000-0002-8716-0482]{Jennifer B. Bergner}
\affiliation{Harvard-Smithsonian Center for Astrophysics, 60 Garden St., Cambridge, MA 02138, USA}

\begin{abstract}
The nature and abundance of sulfur chemistry in protoplanetary disks (PPDs) may impact the sulfur inventory on young planets and therefore their habitability. PPDs also present an interesting test bed for sulfur chemistry models, since each disk present a diverse set of environments. In this context, we present new sulfur molecule observations in PPDs, and new S-disk chemistry models.
With ALMA we observed the CS $5-4$ rotational transition toward five PPDs (DM Tau, DO Tau, CI Tau, LkCa 15, MWC 480), and the CS $6-5$ transition toward three PPDs (LkCa 15, MWC 480 and V4046 Sgr). Across this sample, CS displays a range of radial distributions, from centrally peaked, to gaps and rings. We also present the first detection in PPDs of $^{13}$CS $6-5$ (LkCa 15 and MWC 480), C$^{34}$S $6-5$ (LkCa 15), and \ce{H2CS} $8_{17}-7_{16}$, $9_{19}-8_{18}$ and $9_{18}-8_{17}$ (MWC 480) transitions. Using LTE models to constrain column densities and excitation temperatures, we find that either $^{13}$C and $^{34}$S are enhanced in CS, or CS is optically thick despite its relatively low brightness temperature. Additional lines and higher spatial resolution observations are needed to distinguish between these scenarios. Assuming CS is optically thin, CS column density model predictions reproduce the observations within a factor of a few for both MWC 480 and LkCa 15. However, the model underpredicts \ce{H2CS} by 1-2 orders of magnitude. Finally, comparing the \ce{H2CS}/CS ratio observed toward the MWC~480 disk and toward different ISM sources, we find the closest match with prestellar cores.
\end{abstract}
\keywords{astrochemistry --- ISM: molecules --- methods: observational --- methods: numerical --- protoplanetary disks --- techniques: interferometric}

\section{Introduction} 
\label{sec:intro}

Planets form in dust and gas-rich disks, protoplanetary disks (PPDs), around young stars. Disk chemical structures thus link
interstellar chemistry to the chemistry of nascent planets. The chemical composition of these disks is set by
a combination of inheritance from their parent molecular cloud and 
{\it in situ} processes \citep[e.g.][]{aikawa1999,willacy2007,willacy2009,dutrey2014,sakai2014,cleeves2014,oberg2015nat}. 
Much of this chemical composition is still poorly constrained, however, and in the case of sulfur, its chemistry and main reservoirs are yet largely unknown \citep[\eg][]{semenov2018,phuong2018}.

Sulfur chemistry is not just poorly understood in disks, but also in many other circumstellar and interstellar environments. In the
diffuse interstellar medium (ISM) and in photon-dominated regions
(PDRs) the observed sulfur abundance is close to the cosmic value
\citep[\eg][]{howk2006,neufeld2015,goicoechea2006},
whereas in dense molecular gas it is found highly depleted: only 0.1\%
of the sulfur cosmic abundance is observed in the gas phase \citep{tieftrunk1994}, implying a depletion factor of three orders of magnitude
\cite[\eg][]{wakelam2004,vastel2018}. This level of depletion suggests that most sulfur is locked into icy mantles coating interstellar dust grains \citep{millar1990,ruffle1999,vidal2017,laas2019}. However, OCS
and tentatively \ce{SO2} are yet the sole molecules detected in icy
grain mantles toward high-mass protostars \citep[\eg][]{geballe1985,palumbo1997,zasowski2009} and their abundances ($\sim
\dix{-7}$) make up less than 4\% of the sulfur cosmic abundance. Icy \ce{H2S},
the most obvious sulfur ice chemistry product from successive
hydrogenations of atomic sulfur in solid phase, has not been detected
yet \citep{smith1991,boogert2015}.

While most of the sulfur is hidden from us in star forming regions, recent spectral line surveys have increased the number of known interstellar sulfur molecules to a dozen of species in
prestellar cores \citep{vastel2018}, protostellar envelopes
\citep{drozdovskaya2018} and PDRs (Riviere-Marichalar et al. submitted). Together with new models, these new detections have advanced our understanding of the ISM gas-phase S-chemistry. 
In particular, we now know that a rich interstellar sulfur chemistry is present in various ISM environments. The
main S-reservoirs have not yet been identified, however, and there is still much theoretical work left to do to identify the chemical pathways that produce the observed distributions of sulfur species. 

It is also unclear how much of the interstellar sulfur molecules survive during star formation and become incorporated into planet-forming disks, or whether the sulfur chemistry in such disks is largely reset. Sulfur species are present in the remnants of our own PPD, comets \citep[e.g.][]{boissier2007,leroy2015} which, could actually even be relics from its parent ISM molecular cloud according to the Rosetta analysis of the 67P/Churyumov-Gerasimenko cometary ices \citep[e.g.][]{altwegg2016}. In theory, the cometary sulfur molecule abundance patterns and isotopic ratios could be used to constrain whether they are of interstellar or disk origin, but this requires a much better understanding of disk sulfur chemistry than is currently available. A first motivation for pursuing disk sulfur studies is therefore to better constrain the chemical origins of the Solar System. A second motivation is to enable predictions of what kind of sulfur molecules become incorporated into young planets, since this may affect their hospitality to origins of life \citep{ranjan2018}. 

A third motivation is that disks provide a diverse set of chemical  environments and could therefore be used to test and develop a holistic sulfur chemistry model.
In particular, protoplanetary disks are vertically stratified into disk atmospheres, warm
molecular layers and cold midplanes that are analogous to PDRs,
molecular clouds and dense cloud cores, respectively. 
Dependent on which region that dominates the observed emission, disks may express PDR, cloud or core-like sulfur chemistries. Which one is observed may vary between disks around T Tauri and Herbig Ae stars, and perhaps even across individual disks.

Sulfur bearing molecules have been detected in PPDs, but the detections are scarce compared to the earlier stages of star formation described above. So far, only three S-bearing species detections have been reported in the literature. CS is detected in dozens of disks \citep[\eg][]{dutrey1997,fuente2010,guilloteau2016,dutrey2017,teague2018,phuong2018}, while SO is only found toward a few young disks presenting active accretion signs, including  the Herbig A0 star AB Aur and Class I sources \citep[\eg][]{fuente2010,guilloteau2013,guilloteau2016,pacheco2016,sakai2016,booth2018}. Finally, very recently, the first detection of \ce{H2S} was reported toward the T Tauri star GG Tau A \citep{phuong2018}. 

Here we present a holistic analysis of new observations of S-species in disks using the Atacama Large Millimeter/submillimeter Array (ALMA). The survey includes the CS $5-4$ rotational transition observed toward five Taurus PPDs (DM Tau, DO Tau, CI Tau, LkCa 15, MWC 480), the CS $6-5$ transition observed toward two of these disks (LkCa 15 and MWC 480) and V4046~Sgr, and the first detection in PPDs of the species $^{13}$CS, C$^{34}$S, and \ce{H2CS} in their $6-5$, and $8_{17}-7_{16}$, $9_{19}-8_{18}$ and $9_{18}-8_{17}$ rotational transitions, respectively. The latter three are observed toward LkCa~15 and/or MWC~480 (see also Loomis et al. in prep.). These observations are compared to two dimensional disk chemistry models. 

The outline of the paper is the following. The observational details are presented in \S\ref{sec:obs} 
and the observational results in \S\ref{sec:obs-results}, including   
derived column densities and excitation temperatures for
CS and \ce{H2CS} toward LkCa~15 and MWC 480. 
To better understand the observed PPD S-chemistry, we ran two disk chemistry models tuned to MWC~480 and LkCa~15, which we present in \S\ref{sec:modeling}. The observational and modeling results are compared and discussed in \S\ref{sec:discussion} and the conclusions presented in \S\ref{sec:conclusion}.

\section{Observations}
\label{sec:obs}

\subsection{Sample}
\label{subsec:obs-sample}
Our disk sample is composed of six PPDs. Five of them -- surrounding the DM~Tau, DO~Tau, CI~Tau, MWC~480, and LkCa~15 stars -- are situated in the nearby Taurus star forming region (from 139 to 169~pc, \citep{gaia_catalog2018}, see Table~\ref{tab:disk-params}). Their ages range from $\sim1$ to $\sim 10$~Myr (see Table~\ref{tab:disk-params}), and host large ($>200$~au) gas-rich disks \citep[e.g.][]{chiang2001,pietu2007,oberg2010,isella2012,guilloteau2016,huang2017}. 

The sixth PPD is orbiting V4046~Sgr, an isolated binary system located in the $~23\pm3$Myr old $\beta$-Pictoris Moving Group \citep{mamajek2014}.
Among the six disks, three are known to be transitional disks, namely DM~Tau, LkCa~15 and V4046 Sgr, meaning that they present large gaps and/or inner cavities in their submillimeter/millimeter dust continuum emission \citep{andrews2009,pietu2006,rosenfeld2013}.

\subsection{Observational details}
\label{subsec:obs-details}

The observed species transitions, their frequencies and spectroscopic parameters are listed in Table~\ref{tab:obs-list}.

The CS $5-4$ transition was observed in the five Taurus disks (ALMA project code 2016.1.00627.S, PI: K. \"Oberg), in Band 6 during ALMA cycle 4, with a spectral resolution of 61~kHz. These observations were performed in two execution blocks on December 1, 2016 with an angular resolution of $\sim 0.5''$. 
The total on-source integration times were 22-24 min. 44 antennas were included and covered baselines from 15 to 704~m. The first block used the source J0510+1800 as its band-pass calibrator, while the second block used J0237+2848. Both blocks used the J0423-0120 and J0426+2327 source as a flux calibrator and phase calibrator, respectively.

The three \ce{H2CS} $8_{17}-7_{16}$, $9_{19}-8_{18}$ and $9_{18}-8_{17}$ rotational transitions, CS $6-5$ and corresponding $^{13}$CS and C$^{34}$S isotopologues were observed in Band 7 during ALMA cycles 3 \& 4 (project code 2013.1.01070.S, PI: K. \"Oberg), toward two of the Taurus disks: the Herbig Ae disk MWC~480, and the T Tauri disk LkCa~15. The spectral resolution for these lines was $\sim 975$~kHz, i.e. corresponding to a velocity resolution of $\sim 1$~km/s. 

The $^{12}$CS $6-5$ transition was observed in three execution blocks on January 17, 2016, April 23, 2016 and December 12, 2016. For the first execution block 31 antennas were included and covered baselines from 15 to 331~m. 36 antennas were included for the two other execution blocks, covering baselines from 15 to 463~m and 15 to 650~m, respectively. The first and third blocks used the source J0510+1800 as band-pass and flux calibrators, and the source J0438+3004 as phase calibrator. The second block of execution used the source J0238+1636 as band-pass calibrator, the source J0433+2905 as phase calibrator, and the source J0510+1800 as flux calibrator. The total on-source integration time were 17.6, 12.6 and 20.7 min., respectively. 

The $^{13}$CS and C$^{34}$S $6-5$, and \ce{H2CS} $8_{17}-7_{16}$  rotational transitions were observed in a single execution block, on January 17, 2016, with the same calibrator sources as those used for the first and second execution blocks used for $^{12}$CS $6-5$ (see above) but with 36 antennas and a total on-source integration time of $\sim 19.2$ min.

The \ce{H2CS} $9_{19}-8_{18}$ rotational transition was observed in two execution blocks, on December 17 and 18, 2016, with the same calibrator sources as those used for the first and second execution blocks used for $^{12}$CS $6-5$ (see above). 40 and 42 antennas were used with baseline coverages from 15 to 460~m and 15 to 492~m, respectively. The total on-source integration times were $\sim 13.1$ min for each block.

The \ce{H2CS} $9_{18}-8_{17}$ rotational transition was observed in two execution blocks, on December 13 and 14, 2016, with the same calibrator sources as those used for the first and second execution blocks used for $^{12}$CS $6-5$ (see above). 36 and 39 antennas were used with baseline coverages of 15 to 650~m and 15 to 460~m, respectively. The total on-source integration times were $\sim 12.6$ min for each block.

More details on the ALMA Band 7 observations presented above can be found in Loomis et al. in prep.

Finally, the CS $6-5$ transition was observed toward V4046~Sgr as part of the ALMA project 2016.1.01046.S (PI: R. Loomis). These observations were performed in a single execution block on May 06, 2018 with an angular resolution of $0.54''$ and a spectral resolution of 0.1 km/s. The total on-source integration times was 29 min. 44 antennas were included and covered baselines from 15 to 500~m. The quasar J1924-2914 was used as a band-pass and flux calibrator, and J1802-3940 was used as a phase calibrator.

\begin{table*}
    \footnotesize
    \caption{Disk characteristics.}
    \renewcommand{\arraystretch}{1.2}
    \begin{tabular}{lcccccccccc}
    \hline
    \hline
    Source & Stellar Type & R.A.$^{(a)}$&Dec.$^{(a)}$&Dist.$^{(a)}$&$L_\star$&$M_\star$&Disk inc.$^{b}$&Disk P.A.$^{b}$&Age&V$_{\rm{LSR}}$\\
    &  & (J2000) & (J2000)&(pc)&($L_\odot$)&($M_\odot$)&(\degr)&(\degr)&(Myr)&(km/s)\\
    \hline
      DM~Tau & M1-M2$^{[1,2]}$ & 04:33:48.7 & 18:10:10.0& 145&0.2$^{[3]}$&0.5-0.6$^{[3,4]}$&54.4&-25.8&3.5-8$^{[3,5]}$&6.05$^{[6]}$\\
      DO~Tau & M0-M1$^{[1,2]}$ & 04:38:28.6 & 26:10:49.5 & 139& 1.4$^{[3]}$ & 0.5$^{[3]}$ & 30.3 &5.2 &0.8$^{[3]}$  &6.5$^{[6]}$\\
      CI~Tau & K4-K7$^{[1,2]}$ & 04:33:52.0 & 22:50:30.1 & 159 & 0.9$^{[3]}$ & 0.8$^{[3,5]}$ & 50.3 & 194.5&1.5-2.5$^{[3,5]}$ &5.77$^{[6]}$\\
    MWC~480& A1-A3/4 $^{[7,2]}$& 04:58:46.3 & 29:50:37.0 &  162& 19-24$^{[8,3]}$ & 1.7-2.3$^{[3,4,8]}$& 35.3 &-37.0  &6-7.1$^{[3,4,8]}$ & 5.1$^{[9]}$\\
    LkCa~15 & K3-K5$^{[1,2]}$  & 04:39:17.8 & 22:21:03.4 &  159& 0.8$^{[3]}$ & 1.0$^{[3,4]}$ & 49.7 & 61.4&3-5$^{[3,4,5]}$ & 6.3$^{[9]}$\\
    V4046~Sgr$^{(c)}$ & K5, K7$^{[10]}$ & 18:14:10.5 & -32:47:35.3 & 72 &0.5, 0.3$^{[10]}$  & 0.9, 0.9$^{[11,12]}$ & 40.3 &255.5 &  4-13$^{[10,13]}$& 2.9$^{[9]}$\\
    \hline
    \end{tabular}
    \label{tab:disk-params}
    \tablenotetext{a}{Right ascension (R.A.), declination and distance of each source are from the Gaia DR2 catalog \citep{gaia_catalog2018}}
    \tablenotetext{b}{The disk's position angle (P.A.) and inclination angle (inc.), have been derived from an iterative search algorithm that find a pair of P.A. and inc. that best encompassed the disk's CO emission as detailed in Appendix of Pegues et al. in prep.}
    \tablenotetext{c}{V4046~Sgr is a binary star system \citep{quast2000}}
    \tablecomments{ [1]~\cite{herbig_bell1988}, [2]~\cite{luhman2010}, [3]~\cite{andrews2013}, [4]~\cite{simon2000}, [5]~\cite{guilloteau2014}, [6]~\cite{guilloteau2016}, [7]~\cite{the1994}, [8]~\cite{mannings1997}, [9]~\cite{huang2017}, [10]~\cite{quast2000}, [11]~\cite{stempels2004}, [12]~\cite{rosenfeld2013}, [13]~\cite{rosenfeld2012}.}
\end{table*}

\begin{table*}
\footnotesize
\begin{center}
\caption{List of Observations (data from CDMS$^{(a)}$) \label{tab:obs-list}}
 \renewcommand{\arraystretch}{1.2}
\begin{tabular}{lcccclcccccc}
\hline\hline
Species&Transition&Frequency&$E_{up}$&S$_{ij}\mu^2$&Source& RMS$_{\rm{chan}}$& \multicolumn{2}{c}{Restored Beam}&$R_{\rm{\sigma}}^{(b)},R_{\rm{max}}$& $S_\nu\Delta_v$($R_{\rm{\sigma}}$)&$S_\nu\Delta_v$($R_{\rm{max}}$)\\
&&(GHz)&(K)&(D$^2$)&&(mJy/beam)&($''\, \times \,''$)&(\degr)&(AU)&(mJy km/s)&(mJy km/s)\\ 
\hline
$^{12}$CS& $5-4$&244.93564 &35.3&19.3& MWC~480&4.0&$0.45\times0.71$&-12.9&150, 400&$153\pm6$&$536\pm19$\\
&&&&& &&$1.05\times1.26$&-15.9&&$119\pm6$&$509\pm25$\\
&&&&& LkCa~15&2.8&$0.47\times0.57$&20.8&200, 600&$301\pm5$&$1050\pm14$\\
&&&&& &&$1.01 \times 1.15$&37.6&&$233\pm9$&$1065\pm28$\\
&&&&& CI~Tau&2.8&$0.46\times0.57$&16.0&200, 600&$182 \pm 6$&$756 \pm 17$\\
&&&&& DO~Tau&3.1&$0.45\times0.59$&16.7&100, 200&$60 \pm 3$&$159 \pm 6$\\
&&&&& DM~Tau&3.1&$0.47\times0.54$&20.4&200, 500&$159 \pm 5$&$388 \pm 9$\\

\hline
& $6-5$ &293.91224& 49.4&23.1&MWC~480&8.8&$0.36\times0.78$&-26.5&150, 400&$185\pm6$&$688\pm13$\\
&&&&& &&$0.98\times1.28$&-36.6&&$143\pm3$&$628\pm12$\\
&&&&& LkCa~15&8.3&$0.38\times0.70$&-32.0&200, 600&$415\pm10$&$1282\pm25$\\
&&&&& &&$0.96\times1.15$&-47.1&&$319\pm5$&$1339\pm17$\\
&&&&& V4046~Sgr&&&&--, 200&--&$886 \pm 11$\\
\hline
$^{13}$CS& $6-5$ &277.45548& 49.6&23.1&MWC~480&11.5&$1.05\times1.28$&-10.7&150, 400&$\lesssim4$&$\lesssim 6$\\
&&&&& LkCa~15&13.2&$1.02\times1.15$&-43.3&200, 600&$17 \pm 4$&$71 \pm 17$\\
\hline
C$^{34}$S& $6-5$ &289.20923& 38.2&22.3&MWC~480&13.5&$1.01\times1.22$&-11.1&150, 400&$\lesssim6$ &$21 \pm 12$\\
&&&&& LkCa~15&15.2&$0.98\times1.11$&-45.7&200, 600&$24 \pm 4$&$87 \pm 17$\\
\hline
H$_2$CS& $8_{17}-7_{16}$&278.88766&73.4&64.1&MWC~480&13.8&$1.04\times1.28$&-10.7&--, 300&--&$52 \pm 9 $\\
&&&&& LkCa~15&16.5&$1.02\times1.15$&-43.3&--, 320&--& $\lesssim 16$\\
\hline
& $9_{19}-8_{18}$&304.30770&86.2&72.3&MWC~480&8.8&$0.62\times0.94$&-1.0&--, 300&--&$49 \pm 10 $\\
&&&&& LkCa~15&8.8&$0.63\times0.82$&-8.4&--, 320&--& $\lesssim 22$\\
\hline
& $9_{18}-8_{17}$&313.71665&88.5&72.3&MWC~480&15.8&$0.53\times0.95$&-24.8&--, 300&--&$51 \pm 20 $\\
&&&&& LkCa~15&18.3&$0.53\times0.86$&-33.2&--, 320&--&$\lesssim 50$\\
\hline
\end{tabular}
\end{center}
\tablenotetext{a}{\texttt{https://cdms.astro.uni-koeln.de/cdms/portal/}, \cite{cdms2001,cdms2005}}
\tablenotetext{b}{$R_{\rm{\sigma}}$ corresponds to the radius at which the line flux is $\sim 33$\% of the maximum line emission peak, thus focusing on the disk from where most of the CS emission (i.e. 2/3) originates when deriving a characteristic CS column density.}
\end{table*}

\subsection{Data reduction}
\label{subsec:data-reduction}

Data calibration was initially performed by the ALMA/NAASC staff. We then use the {\it Common Astronomy Software Application} package (\texttt{CASA}) version 4.7.2 \citep{mcmullin2007} to reduce the data. Self-calibrations were performed using the continuum emission from each spectral window containing the targeted line.

After continuum subtraction with the CASA {\texttt uvconsub} function, the data were CLEANed \citep{hogbom1974} using a $3 \sigma$ noise threshold and Briggs weighting with a robustness parameter of 0.5 for all the lines. The one exception was the high signal to noise ratio CS $6-5$ line observed toward V4046~Sgr, for which a value of 0 was adopted in order to improve the angular resolution. The RMS per channel are listed in Table~\ref{tab:obs-list}. The masks used for the CLEANing procedure were drawn manually to follow the shape of the emission in each velocity channel.

We applied a Keplerian mask \citep[e.g.][and see details in Appendix of Pegues et al. in prep for our methodology]{salinas2017} to the CLEANed data cubes to improve the signal-to-noise of extracted intensity maps, spectra, and integrated flux for each line. Briefly, the Keplerian mask selects the disk area in which the line emits within each channel, given the beam size, velocity resolution, stellar mass, and disk geometry, and assuming that the disk is 
flat. In theory the flat disk assumption could result in emission that is coming from the flaring layers being cut out. However we pad the mask sufficiently to avoid this and have validated it against CO emission (Pegues et al. in prep.), which is originating in at least as high layers as our molecules due to high optical depths. The Keplerian masks calculated for the different lines studied here are
depicted on the channel maps in Appendix Fig.~\ref{fig:channel-maps-mwc480-cs54}-\ref{fig:channel-maps-mwc480-h2cs_919_818}. The disk and star parameters used to build the Keplerian masks are listed Table~\ref{tab:disk-params}.

\section{Results}
\label{sec:obs-results}

\begin{figure*}
\centering
\includegraphics[scale=0.81]{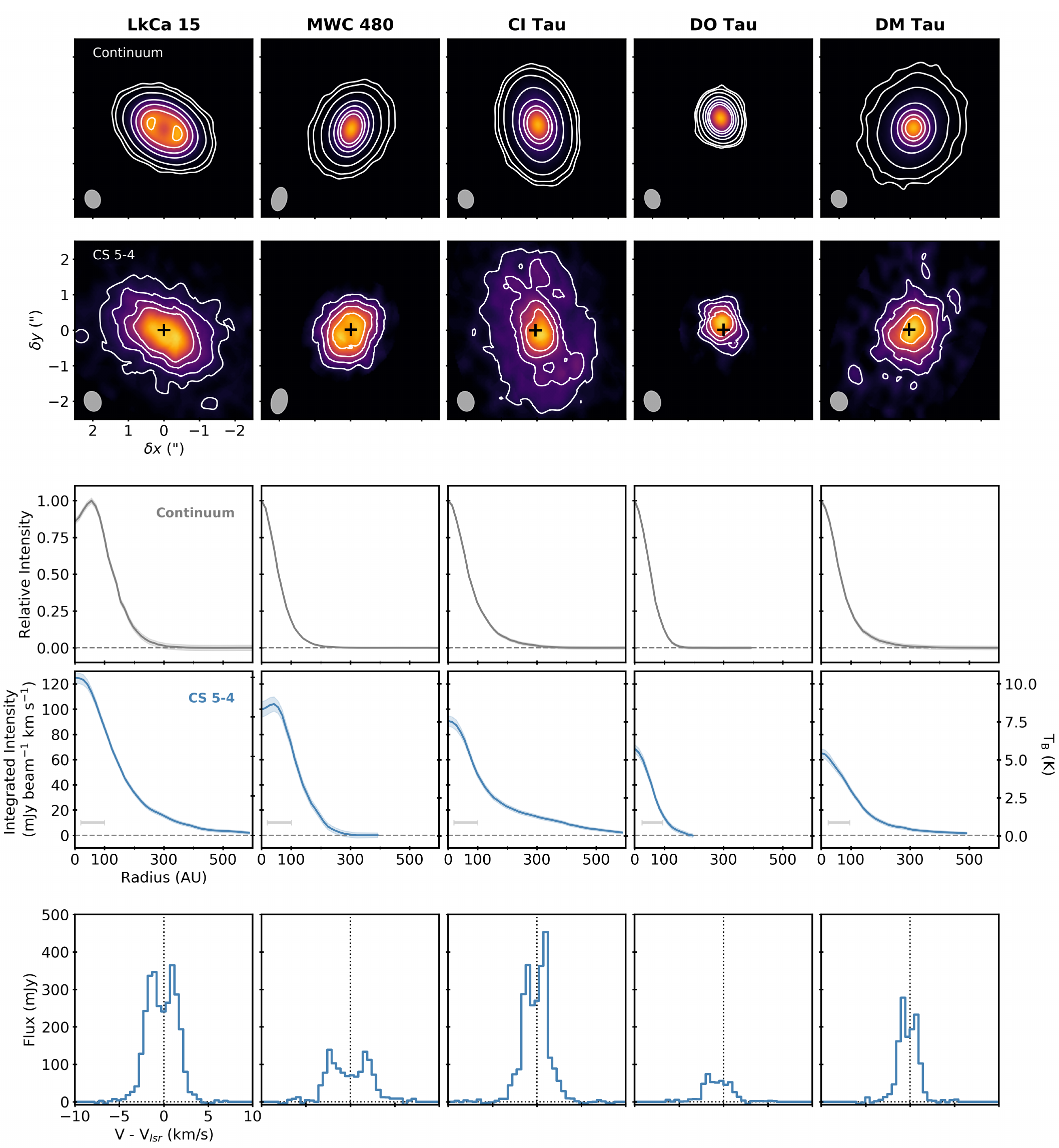}
\caption{ALMA Band 6 observation results toward the five Taurus disks of our sample (each column represents the result for one of the sources). {\it First row:} Integrated intensity maps of the dust continuum at 1.3mm with contours depicted at [5, 10, 30, 100, 200, 350, 500]$\times$ the median RMS. {\it Second row:} Integrated intensity maps of
    CS $5-4$ with contours depicted at [3, 5, 10, 15, 30]$\times$  the median RMS. Synthesized beams are shown in the
    lower left of each panel. The black crosses indicate the center of the continuum.
    {\it Third row:} Radially deprojected and azimuthally averaged flux profiles of the continuum. Beam sizes are represented by the lightgray lines on the bottom left of each panel. {\it Fourth row:} Radially deprojected and azimuthally averaged flux profiles of the CS $5-4$ emission. Note the double y-axis, with on the left the integrated intensity in mJy.beam$^{-1}$.km.s$^{-1}$ units and on the right in Kelvin units. {\it Fifth row:} Integrated intensity spectra of CS $5-4$.  \label{fig:global-fig-obs-CS_5-4}}
\end{figure*}

\subsection{CS fluxes and emission morphologies}
\label{subsec:obs-results-cs}

\begin{figure}
\begin{center}
\includegraphics[scale=0.5]{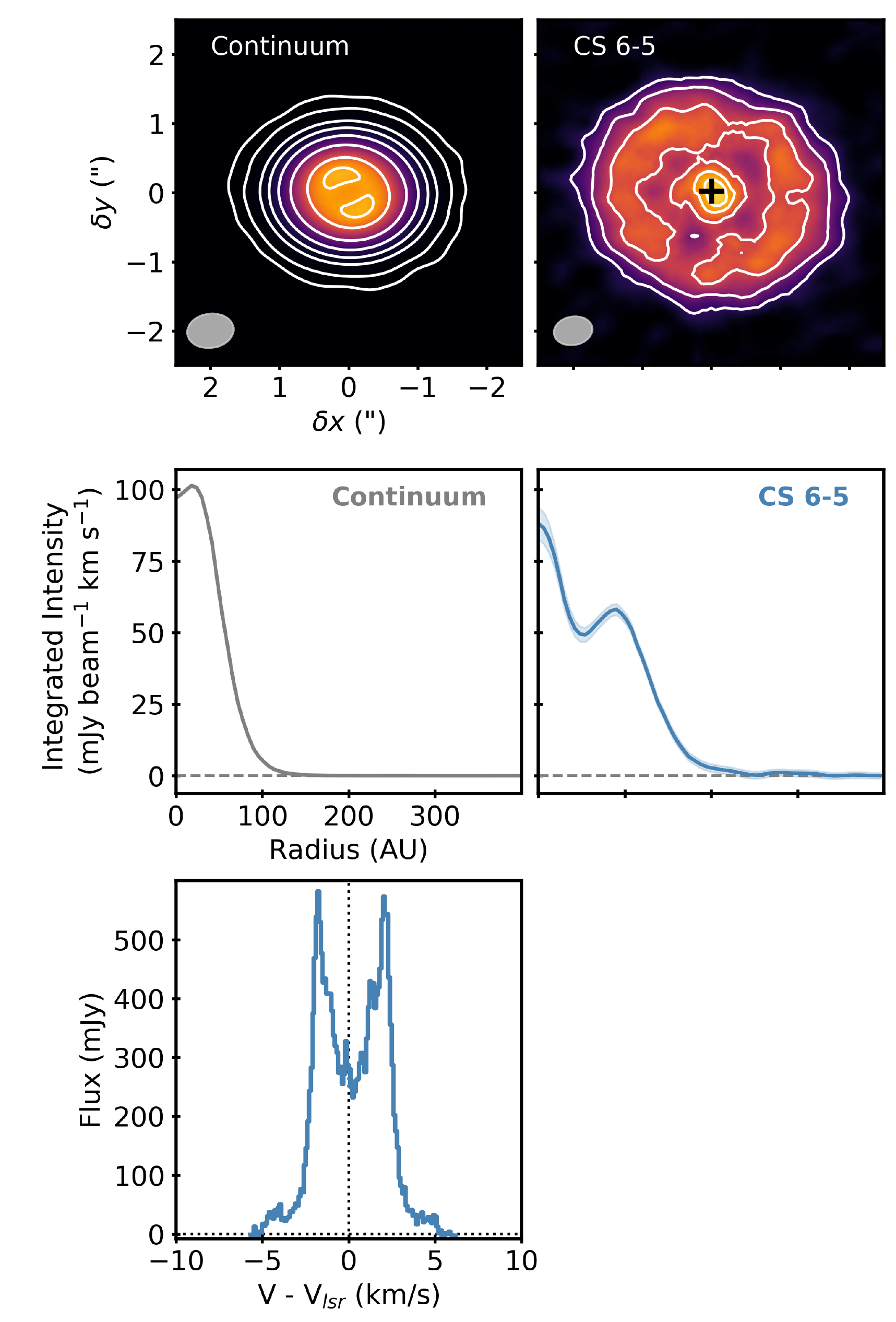}
\caption{ALMA Band 7 results toward V4046 Sgr. {\it First row:} Integrated intensity maps of the 1mm dust continuum and the emission of $^{12}$CS $6-5$, with contour levels shown in white at [10, 30, 100, 200, 400, 800, 1500, 1900]$\times$ the median RMS and [3, 6, 9, 12, 15]$\times$ the median RMS, respectively. Synthesized beams are shown in the lower left of each panel. {\it Top right panel:} The black cross indicates the center of the continuum. {\it Second row:} Radially deprojected and azimuthally averaged flux profiles of the continuum and the $^{12}$CS $6-5$ emission. {\it Third row:} Integrated intensity spectra of $^{12}$CS $6-5$ emission. \label{fig:v4046sgr-global-fig}}
\end{center}
\end{figure}

 Figure~\ref{fig:global-fig-obs-CS_5-4} shows the spatially and spectrally resolved CS $5-4$ line detections toward the five Taurus disks of our sample, together with their corresponding $\sim1.3$~mm continuum emission maps. The angular resolution is $\sim 0.5''$, which corresponds to $\sim 80$ au. The spectra of the CS $5-4$ are also depicted in Fig.~\ref{fig:global-fig-obs-CS_5-4}. For four of our targets, a typical double-peaked profile tracing the Keplerian rotation of the disk, while the double-peaked signature is not obvious toward the faintest, DO~Tau, PPD. This could be due to spectral resolution issue, but DO~Tau is supposed to be very young so there may also be a non-Keplerian component. Further DO Tau studies would help in better constrain its physical structure. Figure~\ref{fig:v4046sgr-global-fig} shows CS $6-5$ line and the $\sim1$~mm continuum emission maps toward V4046~Sgr at an angular resolution of $\sim 0.5''$, which corresponds to $\sim 32$~au (see source distance in Table~\ref{tab:disk-params}). As shown in the last panel of the figure, the spectra presents typical double-peaked profile corresponding to Keplerian rotation.

Across the sample the CS radial emission profile is centrally peaked in 5/6 disks; the only exception is the disk around Herbig Ae star MWC~480, where a small emission dip is seen at the disk center. In the outer disk the CS emission morphology appears related to the dust emission outer edge. 
In the radial profiles, depicted on the second and third row panels in Fig.~\ref{fig:global-fig-obs-CS_5-4}, there is a break in the CS slope close to the continuum edge in all disks. 
In the V4046~Sgr disk, the relationship between the continuum disk and the CS morphology is even more pronounced; as there is a ring of CS gas at $\sim100$~au, at the edge of the dust continuum (see Fig.~\ref{fig:v4046sgr-global-fig}). 

The spectral profiles also encode information about the CS distributions in the observed disks. All disk spectra present substantial wings. This is again most pronounced toward V4046~Sgr, where the spectral line looks like a superposition of two spectra, a narrow Keplerian profile that carries most of the flux and traces emission from the outer disk ring, and a broad fainter component that traces emission from the inner disk.

The disk-integrated flux densities, $S_\nu \Delta v$, derived from this work are listed Table~\ref{tab:obs-list}. We report integrated fluxes out to two different radii for each source, first out to the radius, $R_{\sigma}$, where the CS flux is $\sim 33$\% of the maximum CS emission peak (thus focusing on the part of the disk where most of the CS emission originates when deriving a characteristic CS column density, see Figs.~\ref{fig:global-fig-obs-CS_5-4} and \ref{fig:v4046sgr-global-fig}), and second out to the furthest radius, $R_{max}$, where CS emission is detected.

\subsection{CS 6-5 and isotopologues in MWC~480 and LkCa~15 disks}
\label{subsec:cs65-and-iso}

\begin{figure*}
\begin{center}
\includegraphics[scale=0.5]{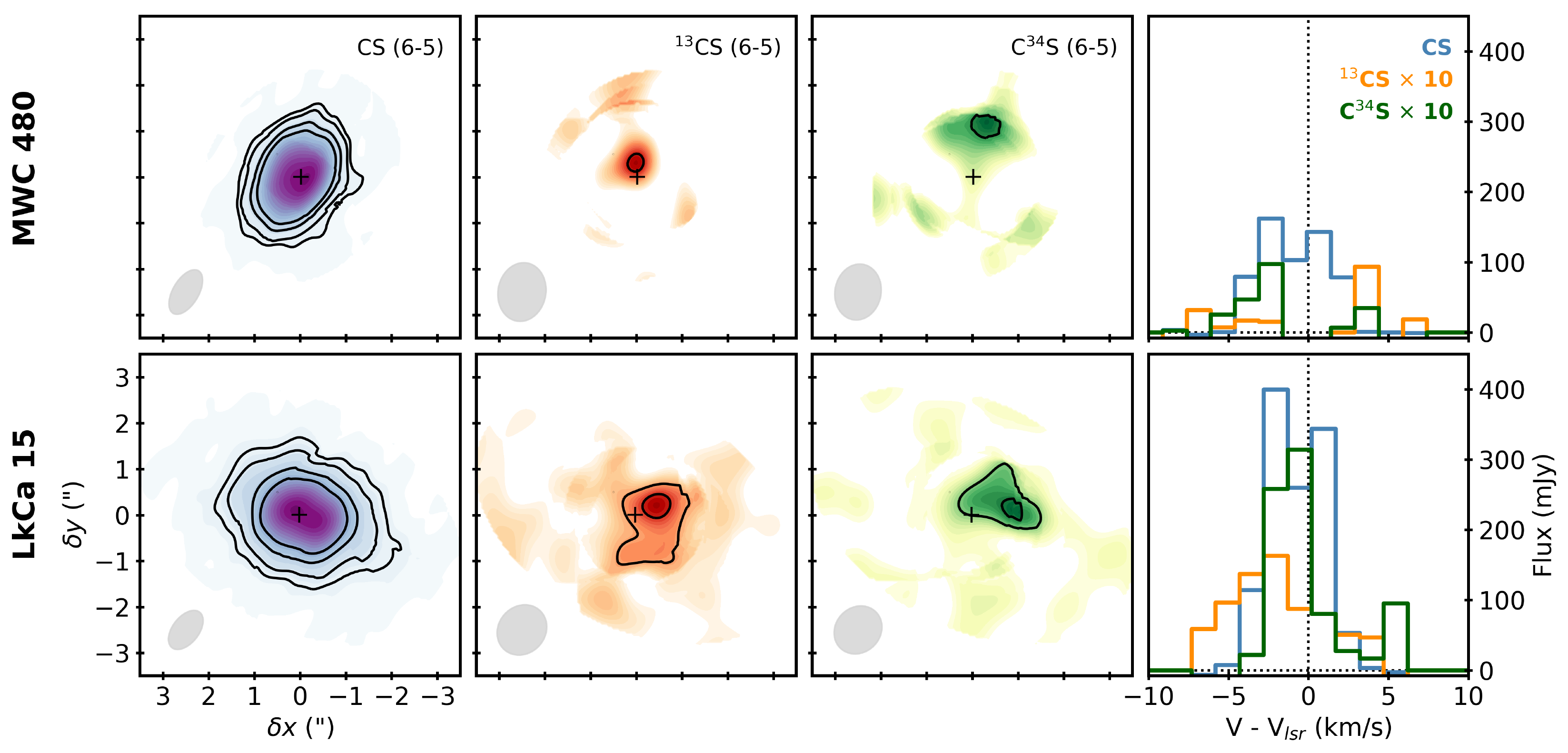}
\caption{Integrated intensity maps and spectra for our
    CS $6-5$ isotopologues observations toward MWC 480 and LkCa 15 in ALMA Band 7. Contour levels are shown at [3, 5; 10; 15]$\times$ the median RMS. Synthesized beams are shown in the lower left of each panel.\label{fig:CS_6-5-mom0maps}}
\end{center}
\end{figure*}

\begin{figure}
\begin{center}
\includegraphics[scale=0.46]{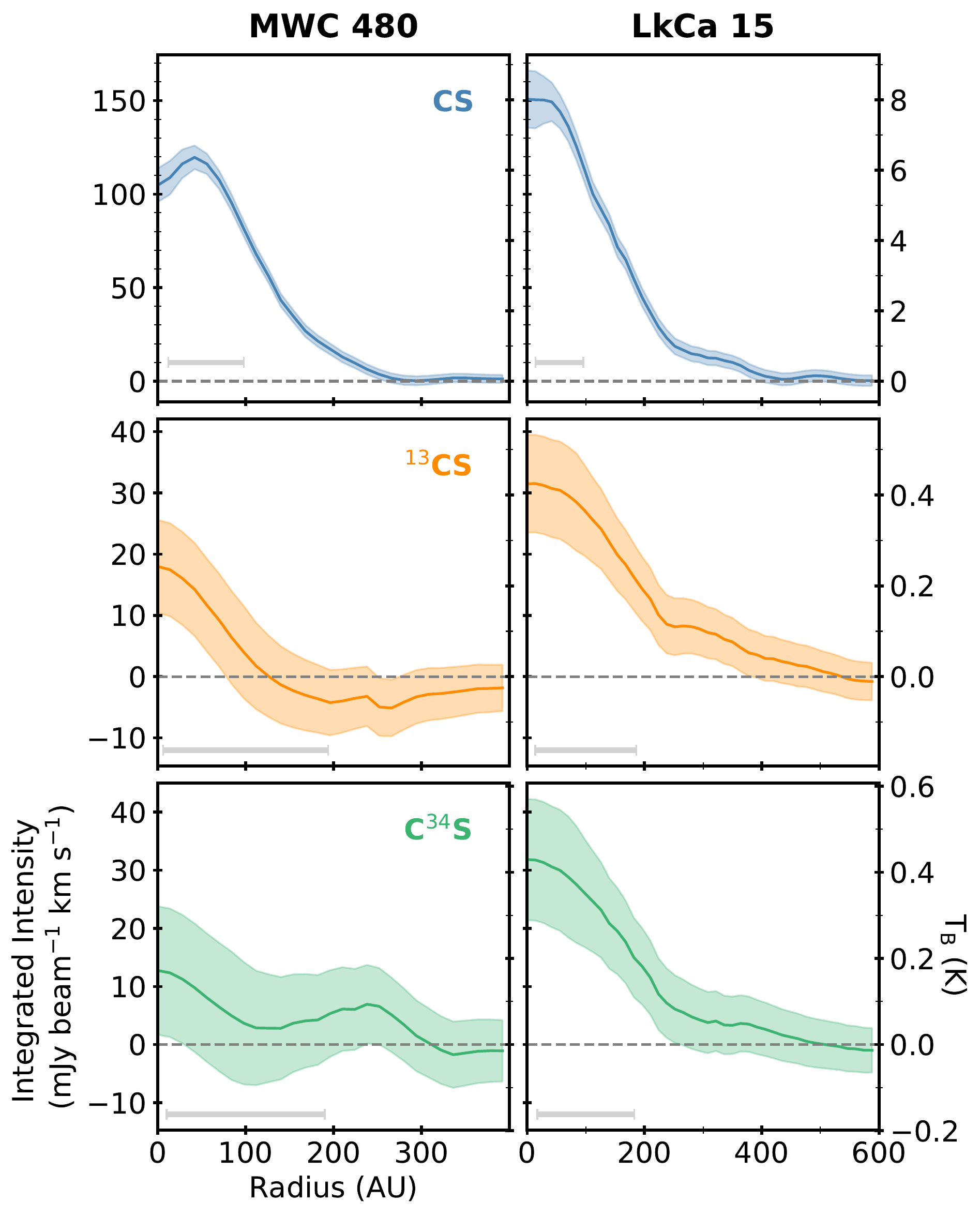}
\caption{Radial profiles of the $6-5$ transitions of $^{12}$CS, $^{13}$CS and C$^{34}$S toward MWC~480 and LkCa~15 respectively. Beam sizes are represented by the lightgray lines on the bottom left of each panel. Note the double y-axis, with on the left the integrated intensity in mJy.beam$^{-1}$.km.s$^{-1}$ units and on the right in Kelvin units. \label{fig:radprof-CS}}
\end{center}
\end{figure}

Toward two disks in the sample, MWC~480 and LkCa~15, three additional CS lines were observed (Fig.~\ref{fig:CS_6-5-mom0maps}). 
The $^{12}$CS $6-5$ line is strongly detected in both disks. 
The peak fluxes are comparable (within 20\%), while the integrated fluxes are somewhat higher compared to the $5-4$ lines
(Table~\ref{tab:obs-list}). 
The $^{12}$CS $6-5$ emission radial morphologies follow those found for $^{12}$CS $5-4$ in the respective disk (Fig.~\ref{fig:radprof-CS}), but the breaks observed at the edge of the continuum are more pronounced.

Figures \ref{fig:CS_6-5-mom0maps} and \ref{fig:radprof-CS} show that the $6-5$ line of the $^{13}$CS and C$^{34}$S isotopologues are detected toward LkCa~15, and tentatively toward MWC~480. The stronger isotopologue emission toward LkCa~15 compared to MWC~480 is consistent with the fact that the former also presents stronger main isotopologue line emission. 

The disk integrated emission of CS, $^{13}$CS and C$^{34}$S $6-5$ in the two disks are reported in Table~\ref{tab:obs-list}. First, whether integrating out to 150/200~au, the radius out to which isotopologue emission is clearly detected, or across the full CS disk, LkCa~15 consistently presents about a factor of two higher flux densities. Second, toward LkCa~15, the flux ratio between $^{13}$CS and the main isotopologue is quite high, which is discussed in detail in \S\ref{subsubsec:iso-ratio}.

\subsection{CS column densities and excitation temperatures}
\label{subsec:coldens-ture-derivation}

Emission line fluxes encode information about the molecular column density and excitation conditions. For the two disks we have observed in two $^{12}$CS lines, we can simultaneously determine the excitation temperature and column density using a rotational diagram analysis \citep[e.g.][]{linke1979,blake1987,goldsmith1999}. This approach implies the simplifying assumptions that the lines are optically thin and at local thermal equilibrium (LTE), and are pursued in \S~\ref{subsubsec:pop-dia}. For one disk, LkCa~15, we have additional information in the form of CS isotopologue detections, and we use these lines to
evaluate the CS column density and excitation temperature in the other limiting case where CS main isotopologue lines are optically thick, in \S~\ref{subsubsec:iso-ratio}. Finally, in \S~\ref{subsubsec:CS_coldens_all}, we use the constraints on the CS rotational temperature from \S~\ref{subsubsec:pop-dia} to estimate the CS column density in all of our disk sample.

\subsubsection{Rotational diagram analysis}
\label{subsubsec:pop-dia}

The first assumption for rotational diagram analysis is that the lines are in LTE. Lines can be assumed to be in LTE when their critical densities are surpassed by the gas densities in their emitting regions. The critical densities of the CS $5-4$ and $6-5$ transitions for temperatures of 20--50~K are $\sim [1.7-1.3]\tdix{6}$ and $\sim [2.9-2.2]\tdix{6}\ccc$ \cite{shirley2015}, respectively. These are lower than expected in molecular disk layers (see e.g. Fig.~\ref{fig:Model-structure}) and LTE thus seems like a reasonable approximation. Under LTE, the population of an energy level $u$ can be approximated as $T_{{\rm ex}}=T_{{\rm kin}}$. The second assumption for rotational diagram analysis is that the lines are optically thin. This assumption can be checked by inspecting the line brightness temperature, which will be substantially below the ambient temperature for optically thin lines.
Fig.~\ref{fig:global-fig-obs-CS_5-4} and \ref{fig:radprof-CS} show that CS brightness temperatures are below 7 and 9~K, for the MWC~480 and LkCa~15 disks respectively. Thermalized lines usually have brightness temperatures which are $\sim 0.6-0.8*T_{{\rm kin}}$ (unless optical depths are extreme), meaning that a 9~K brightness temperature line could partly be optically thick if $T_{{\rm kin}} \sim 11-15$~K. But, in particular for MWC~480, the lowest expected gas temperature in CS emitting regions is $\sim20-30$~K (e.g. \cite{teague2018,semenov2018}, and \S~\ref{sec:modeling}), which suggests that the lines are probably not optically thick. 

Following the LTE and optically thin assumptions, the total column densities and rotational temperatures of CS can be derived from the disk-integrated flux densities, as described in appendix~\ref{app:Nu_SnuDv}. We then used the likelihood function $\mathcal{L}(N_u,\tex)$ described in appendix~\ref{app:Nu_SnuDv} with the Python implementation \texttt{emcee} \citep{emcee2013} of the affine-invariant ensemble sampler for Markov chain Monte Carlo (MCMC) \citep{goodman2010} to fit the data and compute posterior probability distributions for \tex\ and $N_u$ describing the range of total column densities, $N_{\rm {tot}}$, and excitation temperatures, \tex, consistent with the data. The following uninformative priors were assumed:
\begin{eqnarray}
    \tex (\rm{K}) = \mathcal{U}(3,300)\\
    N_{\rm {tot}} (\cc) = \mathcal{U}(\dix{7},\dix{20}).
\end{eqnarray}

\begin{figure}
\begin{center}
\includegraphics[scale=0.9]{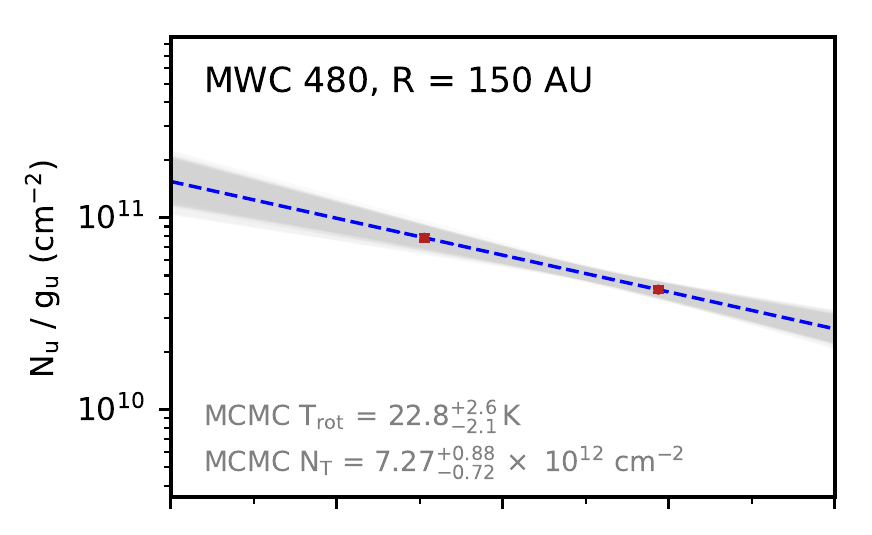}
\includegraphics[scale=0.9]{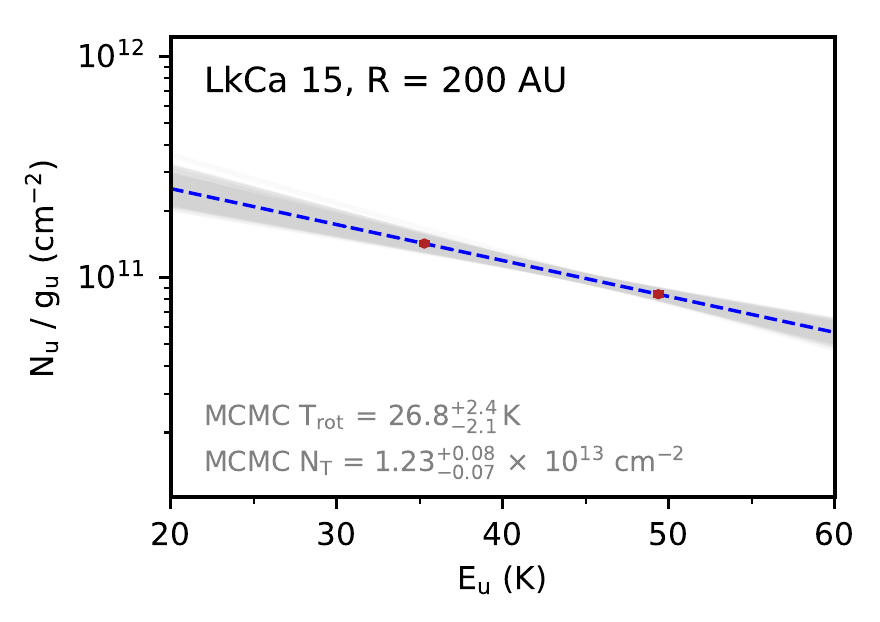}
\caption{CS rotational diagram built from the integrated emission fluxes of the CS $5-4$ and $6-5$ lines, 
integrated over a radius of 150~au toward MWC~480 {\it(top panel)}, and over 200~au toward LkCa~15 {\it(bottom panel)}, with a resolution of $\sim0.5''$ for both.
\label{fig:CS-rot-dia}}
\end{center}
\end{figure}

Random draws and results from the posterior distributions are depicted in gray in Fig.~\ref{fig:CS-rot-dia}, along with the least square fit results, shown in blue. For the MWC~480 disk, the results converged toward \{$\tex\simeq22.8\pm2.5$~K, $N_{\rm {tot}}\simeq 7.3\pm0.9\tdix{12}\cc$\} for a flux integrated over a 150~au radius. For the LkCa~15 disk, the results converged toward
\{$\tex\simeq26.8\pm2.3$~K, $N_{\rm {tot}}\simeq 1.2\pm0.1\tdix{13}\cc$\} for a flux integrated over a 200~au radius. 
The calculated optical depths are $0.07-0.3$ and $0.1-0.4$, toward the MWC~480 and LkCa~15 disks, respectively. While not optically thick the upper boundaries are close enough to $\tau = 1$ to be somewhat affected by opacity. Column densities derived using the optically thin approximation may therefore be underestimated by up to a bit more than an order of magnitude. Higher resolution observations and additional CS are needed to better determine the true CS opacity.

\begin{figure}
\begin{center}
\includegraphics[scale=0.5]{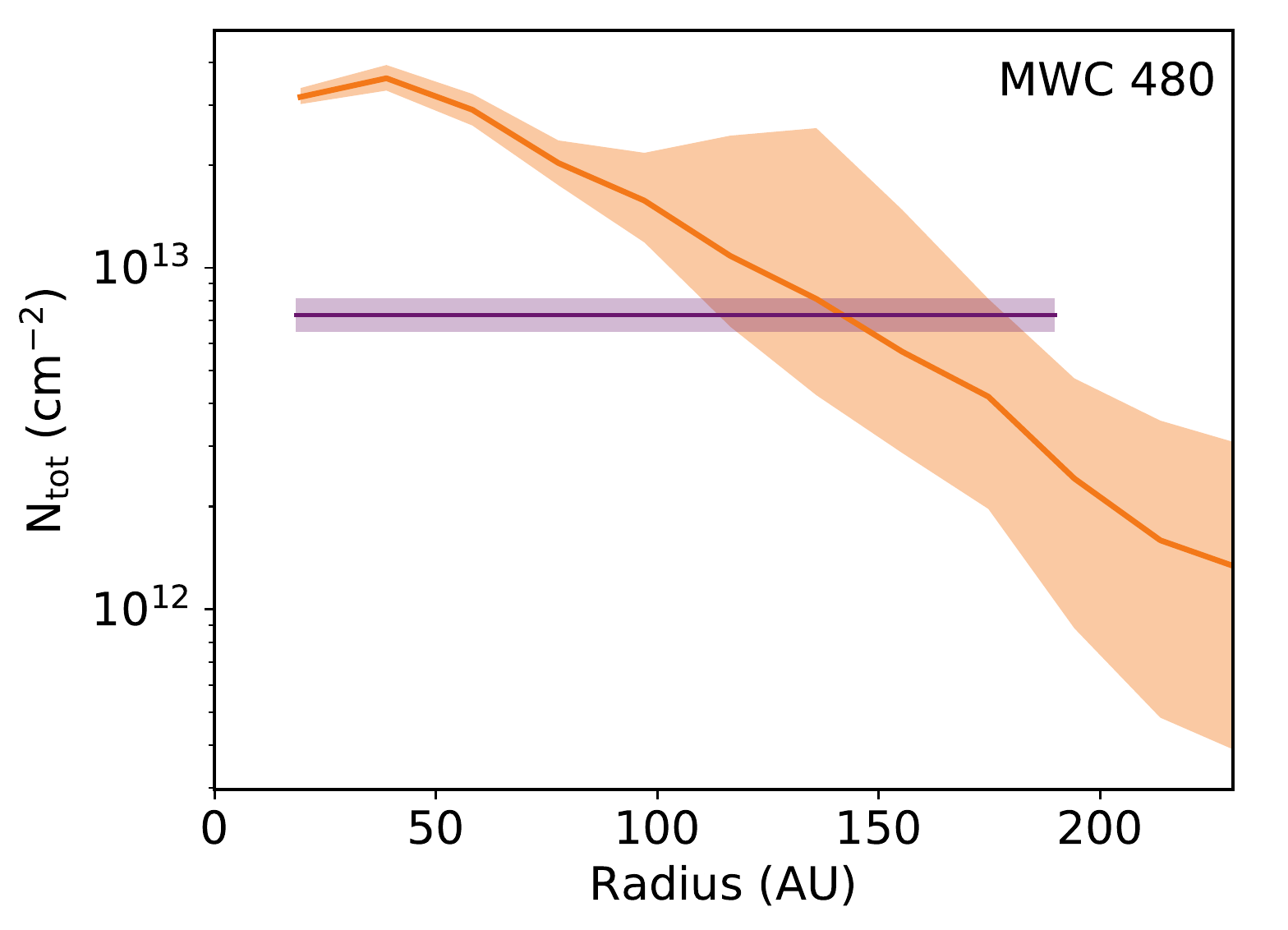}
\includegraphics[scale=0.5]{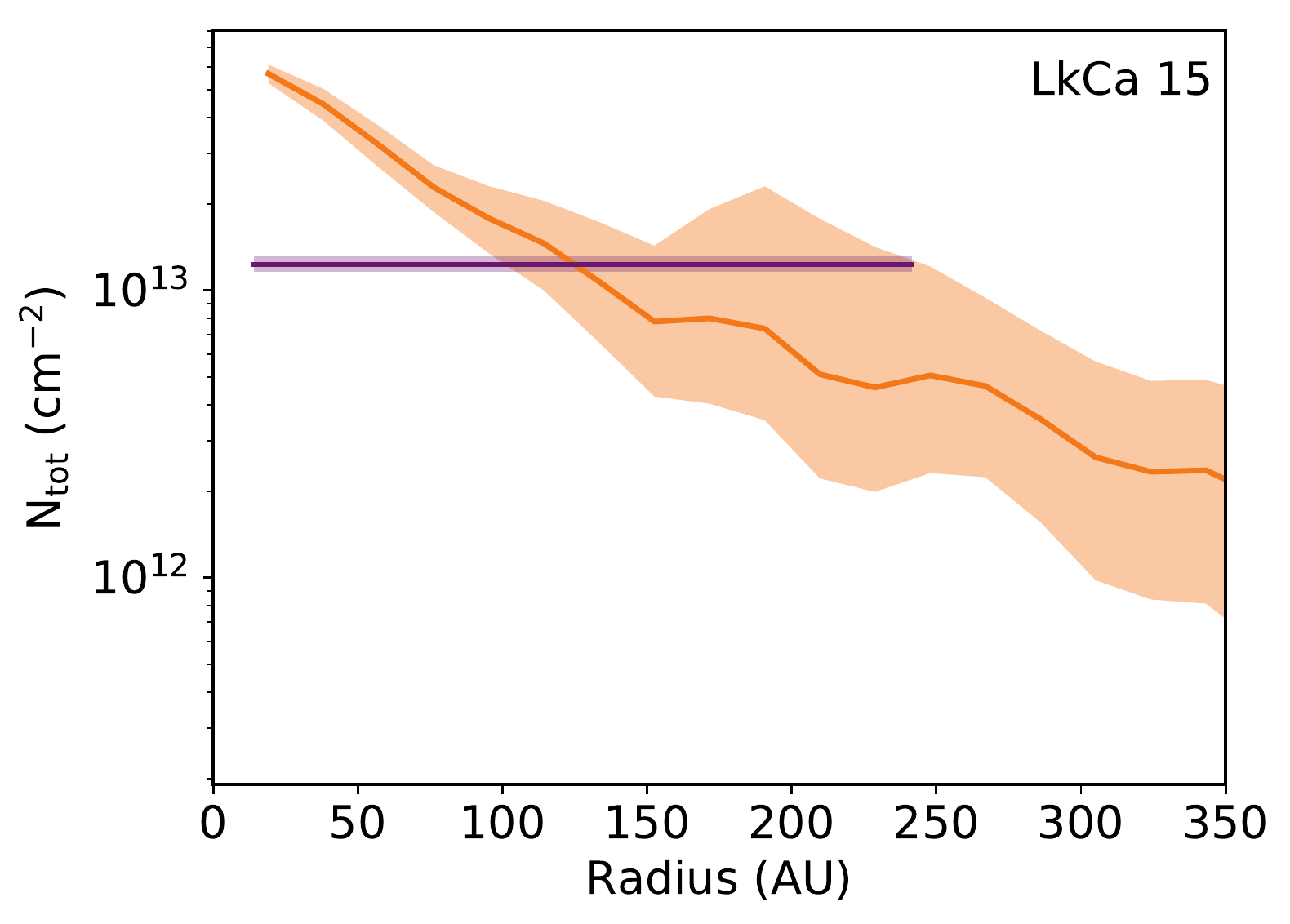}
\caption{Radial profiles of the MCMC rotational diagram results for the column density of CS toward MWC~480 {\it(top panel)} and LkCa~15 {\it(bottom panel)}. Best fit values and 1 $\sigma$ uncertainties are depicted in orange, and the disk-integrated results up to $R_\sigma+0.5\times$beam are overplotted in purple for comparison.
\label{fig:CS-radial-rot-dia}}
\end{center}
\end{figure}

We next calculate the column density of CS as a function of the radius, using the same rotational diagram analysis to the azimuthally deprojected radial intensities.
The results are presented in Fig.~\ref{fig:CS-radial-rot-dia}, showing, as expected, that the CS column density decreases
with increasing radius. We note that in the case of MWC 480, the column density decreases toward the central star, in line with its previously noticed centrally dipped flux profiles.

\subsubsection{Isotopologue ratio analysis}
\label{subsubsec:iso-ratio}
The analysis in \S~\ref{subsubsec:pop-dia} assumed the main isotopologue CS lines are optically thin. This assumption can be explored for one disk, LkCa~15, where two CS isotpopologue lines, $^{13}\ce{CS}$ and C$^{34}$S $6-5$, are well detected.

The ratio of the two isotopologues $^{12}$CS and $^{13}$CS can be expressed as:
\begin{equation}
\frac{T_{B}(\rm{^{13}CS})}{T_{B}(\rm{^{12}CS})} \, = \, \frac{1-e^{-\tau_{\,\rm{^{13}CS}}}}{1-e^{-\tau_{\, \rm{^{12}CS}}}},
\label{eq:iso-ratio}
\end{equation}
 with $T_B$ the brightness temperature (see Appendix~\ref{app:rad-tr-eqn}), and provided that the transitions have the same uniform excitation temperature along the line of sight, and emit over the same volume of gas. The former is approximately true when comparing CS and $^{13}$CS $6-5$, and the latter is a reasonable approximation if the main isotologue line does not become optically thick high up in the disk atmosphere. Furthermore, if $^{12}$CS and $^{13}$CS are both optically thin then we have $(1-e^{-\tau})\rightarrow\tau$ which leads to:
\begin{equation}
\frac{T_{B}(\rm{^{13}CS})}{T_{B}(\rm{^{12}CS})} \, = \,
\frac{\tau_{\rm{^{13}CS}}}{\tau_{\rm{^{12}CS}}}\, \approx\, R.
\label{eq:iso-ratio-thin-approx}
\end{equation}
where $R$ is the abundance ratio of $^{12}$CS to $^{13}$CS. 
Assuming that their abundance ratio is equal to the $^{12}$C/$^{13}$C ratio determined in the local interstellar medium (LISM), i.e. that their has been no isotopic fractionation, $R\simeq 68\pm15$. \citep{milam2005,asplund2009,manfroid2009}.

For optically thin lines we can calculate the disk-integrated brightness temperatures from the disk integrated fluxes $S_\nu \Delta v$
 using the Rayleigh-Jeans law:
\begin{equation}
    T_{\rm{B}} \Delta v = \frac{c^2 S_\nu \Delta v}{2 k_{\rm{B}} \nu^2 \Omega}.
\label{eq:tb}
\end{equation}

To consistently calculate disk integrated fluxes for the main and rare CS isotopologues we applied a {\it uv}-taper to the extended baselines of the $^{12}$CS visibilities before integrating the fluxes, to match the synthesized beams. As can be seen in Table~\ref{tab:obs-list} this reduces the calculated flux by $\sim$25\%. 
For the LkCa~15 disk we then obtain:

\begin{equation}
\frac{T_{B}(\rm{^{12}CS})}{T_{B}(\rm{^{13}CS})} \, \simeq \, 17 \ll \left(\frac{^{12}\ce{C}}{^{13}\ce{C}}\right)_{\rm{LISM}}.
\label{eq:iso-ratio-13cs}
\end{equation}

If the $^{12}$CS/$^{13}$CS abundance ratio is equal to the $^{12}$C/$^{13}$C LISM ratio, Eq.~\ref{eq:iso-ratio-13cs} implies that the $^{12}$CS $6-5$ transition is optically thick.
This implication can be used to calculate the $^{13}$CS column density and the $^{12}\ce{CS}$ excitation temperature. For an optically thick line 
$(1-e^{-\tau_{^{12}\ce{CS}}})\rightarrow 1$. Thus, Eq.~\ref{eq:transfert} directly provides $T_{ex}(\rm{^{12}CS})$ from the peak of its brightness temperature:
\begin{equation}
T_{ex}(\rm{^{12}CS})=\frac{h\nu/k_{\rm{B}}}{\ln\left(1+\frac{h\nu/k_{\rm{B}}}{T_R(\rm{^{12}CS})/f+J_{\nu}(T_{bg})}\right)} \,.
\label{eq:tex_thick}
\end{equation}

Eq.~\ref{eq:tex_thick} gives an excitation temperature of $T_{ex}(\rm{^{12}CS})\simeq 7$~K which, inserted into~\ref{eq:transfert}, gives an opacity of $\tau_{\rm{^{13}CS}} \simeq 0.06$ (this last step assumes that CS and $^{13}$CS have the same excitation temperatures).
Moreover, since we assume that $^{13}$CS is optically thin we have $(1-e^{-\tau_{^{13}\ce{CS}}})\rightarrow \tau_{^{13}\ce{CS}}$, which inserted into Eq.~\ref{eq:iso-ratio} gives:
\begin{equation}
\frac{T_{B}(\rm{^{13}CS})}{T_{B}(\rm{^{12}CS})} \, \approx \, \tau_{\rm{^{13}CS}}.
\label{eq:rare-iso-opacity}
\end{equation}
Eq.~\ref{eq:rare-iso-opacity} gives an opacity consistent with the value obtained from Eq.~\ref{eq:tex_thick} and \ref{eq:transfert}.

Once the opacity and excitation temperature derived, the column density can be computed as follows:
\begin{equation}
\begin{split}
    N_{\rm{tot}}=\frac{3h}{8\pi^3S\mu^2}\frac{Q_{rot}(T_{\rm{ex}})}{g_u}\frac{e^{E_u/k_BT_{ex}}}{e^{h\nu/k_B T_{ex}}-1}\\ \times \frac{1}{J_\nu(T_{ex})-J_\nu(T_{bg})} \int T_R dv \, C_\tau .
\end{split}
\label{eq:CS-Ntot}
\end{equation}
This gives a column density of \ntot($^{13}$CS)$\simeq 7\tdix{12}\cc$.
Assuming that $^{12}$CS is present at the LISM ratio, 
\ntot($^{12}$CS)$\simeq4.8\tdix{14}\cc$, which is significantly higher than the rotational diagram result, by a factor of forty (see Table~\ref{tab:coldens_ture}). 
This discrepancy can be understood when considering that $\tau_{\rm{^{13}CS}} \simeq 0.06$ thus implies that that $\tau_{\rm{^{12}CS}} > 1$ (assuming LISM $\ce{^{12}C}/\ce{^{13}C}$ ratio). We discuss further below whether the LISM assumption and other assumptions implicit or explicit in the analysis are defendable.

A similar computation can be done for the C$^{32}$S/C$^{34}$S ratio considering the elemental 
$^{32}$S/$^{34}$S$= 24.4\pm5.0$ in the vicinity of the Sun \citep{chin1996}. 
Both observed $^{12}$CS($6-5$)/C$^{34}$S($6-5$) and $^{12}$CS($5-4$)/C$^{34}$S($6-5$) give a ratio of $\sim$13, which is somewhat lower than the cosmic $^{32}$S/$^{34}$S ratio, but not nearly as much below the cosmic ratio as $^{12}$CS/$^{13}$CS. Taken at face value, the excitation temperature and opacity are \tex($^{12}$CS)$\simeq 7.4$~K and $\tau_{^{34}\ce{CS}} = 0.07$, leading to \ntot(C$^{34}$S)$\simeq 6\tdix{11}\cc$. Using the LISM ratio we thus obtain \ntot($^{12}$CS)$\simeq 1.5\tdix{13}\cc$, which is an order of magnitude lower than the value derived from the $^{12}$CS/$^{13}$CS ratio, and similar to the rotational diagram result.

The results obtained from the different methods presented in \S\ref{subsubsec:pop-dia} and \S\ref{subsubsec:iso-ratio} to derive CS column densities and excitation temperatures are summarized in Table~\ref{tab:coldens_ture}. Both isotopologue analysis suggest that the $^{12}$CS line is optically thick, but the large difference in derived $^{12}$CS opacities and column densities, as well as the low CS excitation temperatures cast some doubt on the approach.
This makes it important to revisit some of the assumptions that went into the calculations. First, these calculations assumed a beam filling factor of unity. Considering the realization that both dust and gas are highly structured in disks \citep{huang2017,favre2018,andrews2018,huang2018} this is not obviously true. A beam filling factor of 20 to 40\% would increase the excitation temperature to $20-30$~K, in better agreement with our model expectations (see \S~\ref{subsec:model-results}).
Although Figure \ref{fig:CS_6-5-mom0maps} shows apparent asymmetries for the $^{13}$CS and C$^{34}$S emissions, these are likely due to noise in the image which propagates to the 0th moment map. So, as such, there is likely little effect on the extracted flux and thus on the isotopologue ratios. Therefore, it seems reasonable to not expect any clear morphological differences between CS and its isotopologues. However, differences in beam-filling factors of the different isotopologues could impact this analysis. Higher signal-to-noise ratio and higher resolution data are needed to improve this analysis. A second assumption is that LTE conditions hold. This should be true if CS is present in the warm molecular layer of disk midplane, but LTE may not hold if CS is instead emitting from the disk atmosphere. 
 
Third, and perhaps most importantly, $^{13}$CS and C$^{34}$S may not be present at the cosmic ratios. There are many examples of isotopic fractionation in disks and the ISM, including in carbon \citep[e.g.][]{woods_willacy2009,sakai2010,sakai2013,ossenkopf2013,smith2015,yoshida2015,yu2016,huang2017}. For instance, the $^{13}$CS enhanced abundances in disk molecular layers could be explained by excessive $^{13}$C-carrier, such as $^{13}$CO (while $^{12}$CO remains self-shielded). Rapid ion-molecule gas-phase reactions could thus transfer $^{13}$C from $^{13}$CO to hydrocarbons, and then to $^{13}$CS. The degree of fractionation that would be needed to explain the $^{13}$CS emission is quite high, and detailed modeling on carbon fractionation in CS would be needed and warranted to support this scenario. In the meantime we adopt the results from the rotational diagram analysis, which also agree with the C$^{34}$S data within error bars, for model comparisons and calculations of column densities in disks without constraints on the excitation temperature.

\subsubsection{CS column density estimates for the disk sample}
\label{subsubsec:CS_coldens_all}
\begin{figure}
    \centering
    \includegraphics{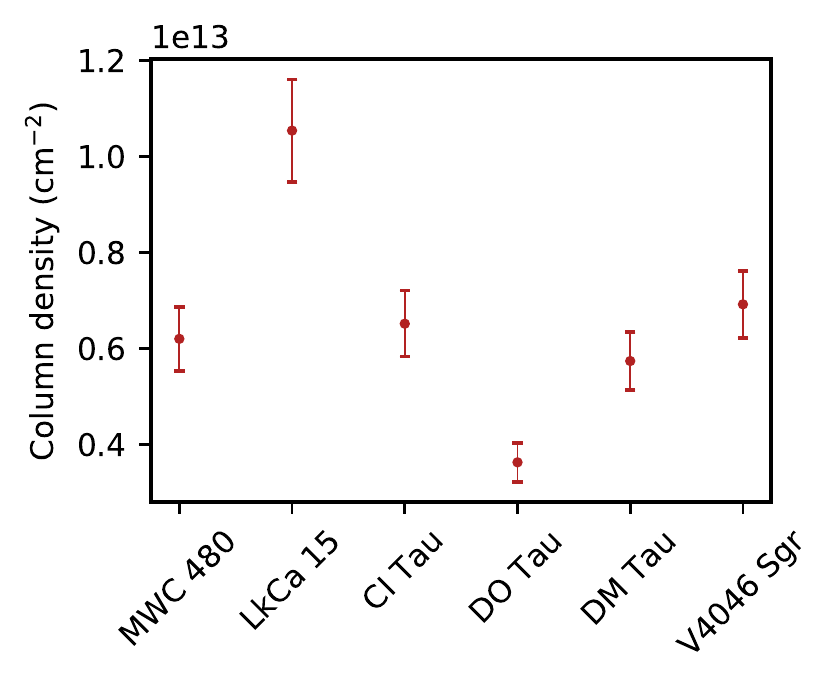}
    \caption{Estimated CS column densities disk-integrated up to $R_\sigma$ for the Taurus disks and $R_{\rm{max}}$ for the V4046~Sgr disk (see Table~\ref{tab:obs-list})
    10\% calibration uncertainty and 20\% uncertainty on the temperature are included.}
    \label{fig:CS_coldens_full_sample}
\end{figure}

In 4/6 disks we observed a single CS line and therefore have no constraint on the CS excitation temperature. To estimate column densities in these sources, we use the constraints on the CS rotational temperature in the LkCa~15 and MWC~480 disks and apply them to the disk sample. In \S~\ref{subsubsec:pop-dia} we found disk averaged rotational temperatures of $\sim23\pm3$, and $\sim27\pm2$~K toward MWC~480 and LkCa~15, respectively. We adopt the average of the two, 25~K, as the typical disk-averaged CS rotational temperature, and an uncertainty of $\pm5$~K. We then use Eqs.~\ref{eq:Nu} and \ref{eq:boltzmann_dist} to calculate the column density.
The results are presented in red in Fig.~\ref{fig:CS_coldens_full_sample}.
The CS disk-integrated column density varies by less than an order of magnitude across the sample of disks. The average column density is $\approx 7\tdix{12}\cc$ and the standard deviation $\approx 2\tdix{11}\cc$.
There are no obvious trends with stellar spectral type; the only Herbig Ae star, MWC 480, is close to the sample average. There is also no trend with disk type, i.e. we see no systematic difference in CS column density between full disks and transition disks. It is interesting that the youngest disk in the sample, DO Tau, presents the lowest value, but apart from this there is also no obvious trend with age as V4046~Sgr, the oldest disk in the sample, lies close to the sample average. In the literature some of those disks are found to be a bit colder \citep[e.g. the DM Tau case,][]{semenov2018}. For colder disk temperatures the derived CS column densities increase, as expected from Eq.~\ref{eq:boltzmann_dist}, leading for instance to an average column density of $\approx 5\tdix{13}\cc$ with a standard deviation of $\approx \pm 5\tdix{12} \cc$ for a disk temperature of $10 \pm 2$~K, i.e. a factor 5 higher column density.

\begin{table*}[]
    \centering
    \caption{Column densities and excitation temperatures derived from the methods presented in Sect.~\ref{sec:obs-results} when applied}
    \renewcommand{\arraystretch}{1.3}
    \begin{tabular}{cccccc}
    \hline\hline
       Species  & Source & Radius & \ntot & \tex & Method\\
       &  & (AU) & (\cc) & (K) & \\
    \hline
       $^{12}\ce{CS}$  & MWC 480& 150 & $(7.3\pm0.8)\tdix{12}$& $22.8\pm2.5$& Rotational diagram\\
     \cline{2-6}
            &LkCa~15 & 200&$(1.2\pm0.1)\tdix{13}$ & $26.8\pm2.3$ & Rotational diagram\\
         & & & $[1.5\tdix{13} - 5\tdix{14}]$& $\sim7$ & Isotopologue ratio analysis\\ 
        \hline
       \ce{H2CS}  & MWC~480 & 300 & $3.0^{+2.7}_{-0.4}\tdix{12}$& $40.5^{+38.4}_{-20.2}$& Rotational diagram \\
            & LkCa~15 & 320 & $\lesssim 2\tdix{12}$& $40$& Using Eqs.~\ref{eq:Nu} and \ref{eq:boltzmann_dist} and a fixed \tex \\
       \hline
    \end{tabular}
    
    \label{tab:coldens_ture}
\end{table*}

\subsection{\ce{H2CS} in MWC~480 and LkCa~15}

\begin{figure*}
\begin{center}
\includegraphics[scale=0.5]{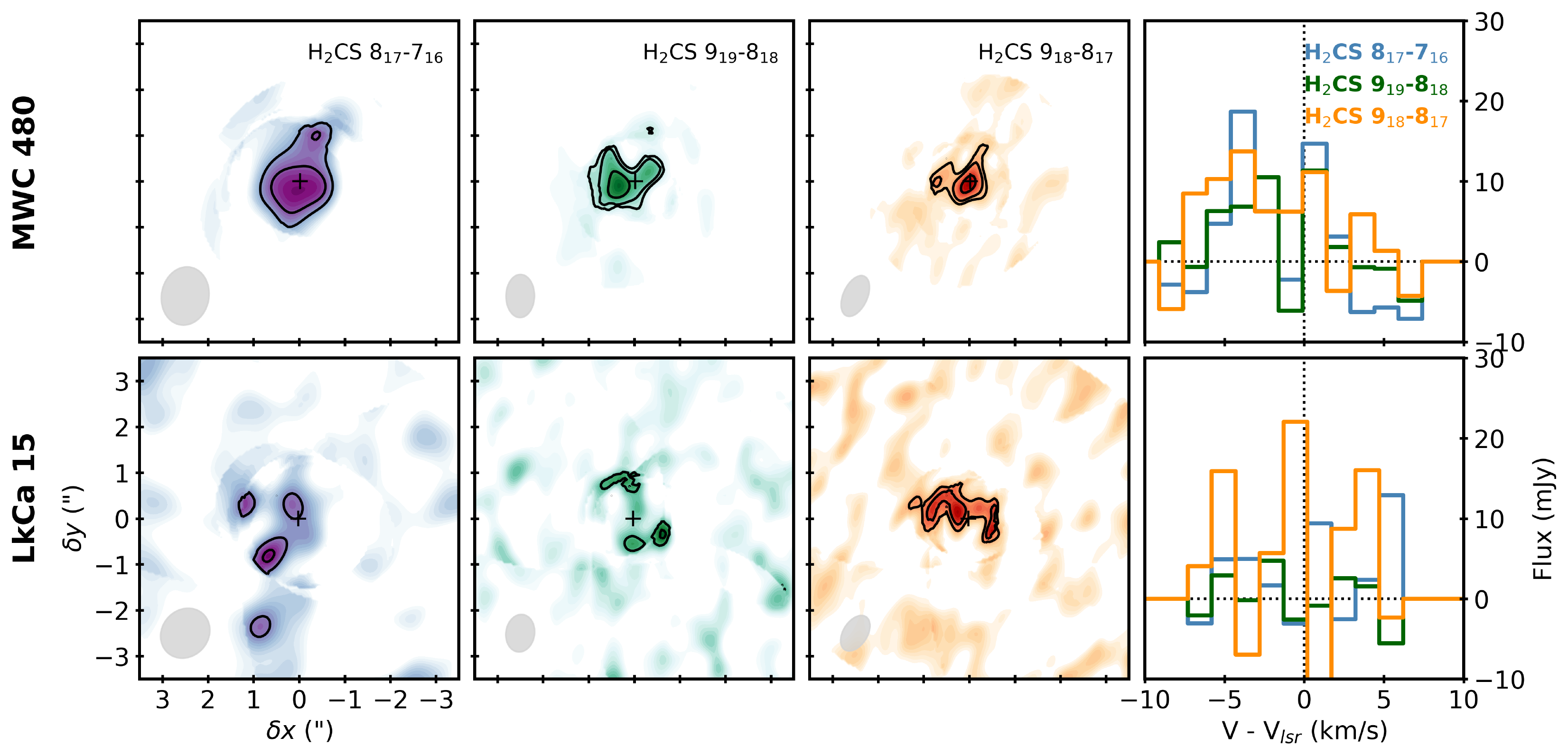}
\caption{Integrated intensity maps and spectra for the three \ce{H2CS} lines
    detected toward MWC~480 and LkCa~15 in ALMA Band 7. Contour levels are shown at [2,3,5]$\times$ the median RMS in black. Synthesized beams are shown in the lower left of each panel in gray.\label{fig:H2CS-mom0maps}}
\end{center}
\end{figure*}

\begin{figure}
\begin{center}
\includegraphics[scale=0.46]{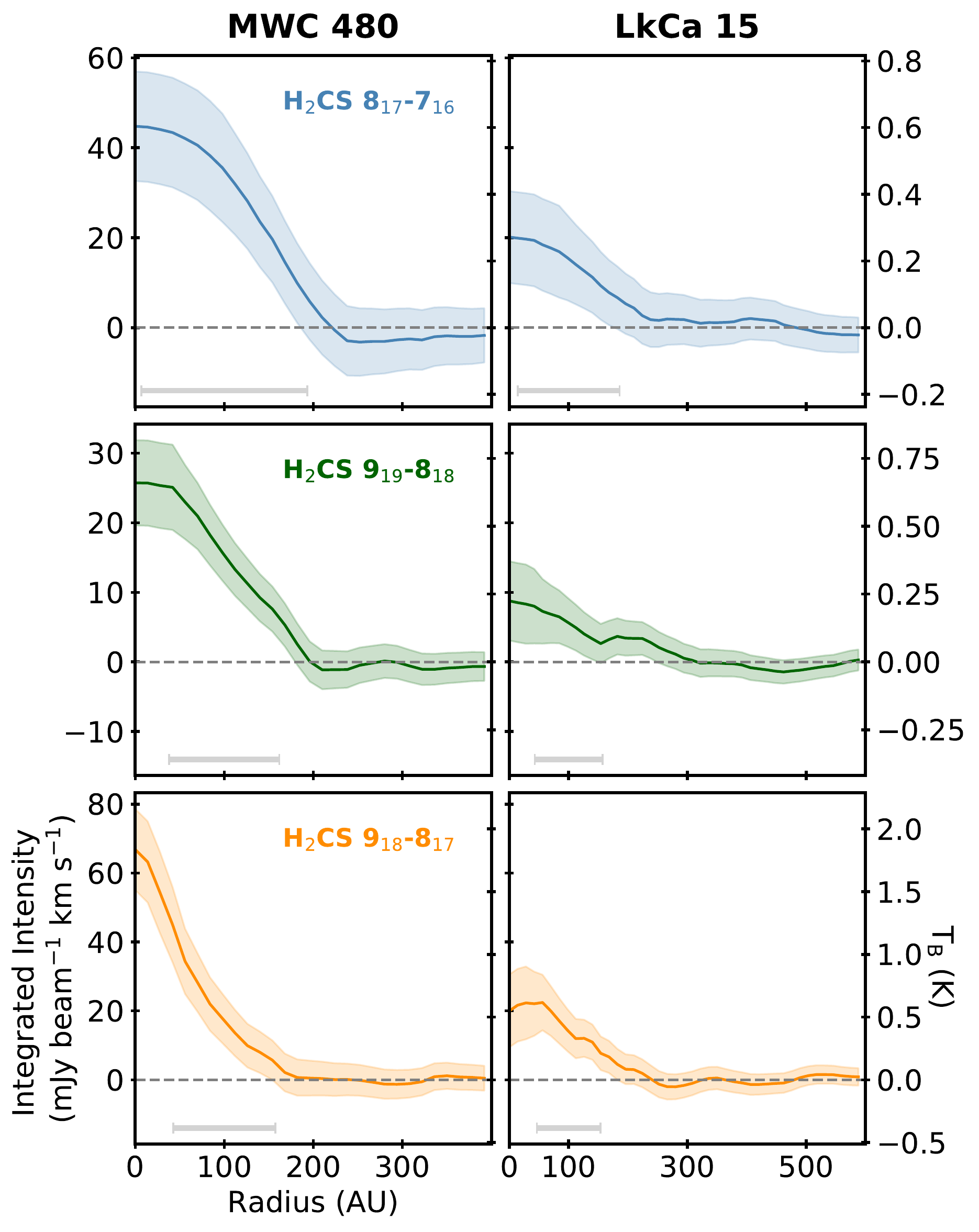}
\caption{ Radial profiles of the three H$_2$CS ortho transitions we have detected toward MWC~480 and LkCa~15 respectively. Beam sizes are represented by the lightgray lines on the bottom left of each panel. Note the double y-axis, with on the left the integrated intensity in mJy.beam$^{-1}$.km.s$^{-1}$ units and on the right in Kelvin units. \label{fig:radprof-H2CS}}
\end{center}
\end{figure}

The three \ce{H2CS} ortho lines ($8_{17}-7_{16}$, $9_{19}-8_{18}$ and $9_{18}-8_{17}$) are detected toward MWC~480 and tentatively detected toward LkCa~15 (2--3$\sigma$). The integrated flux density maps are shown in Fig.~\ref{fig:H2CS-mom0maps} and the radial profiles in Fig.~\ref{fig:radprof-H2CS}. The emission toward MWC~480 shows no sub-structure, including no central depression. However, the resolution and signal-to-noise are too low to rule out a similar central dip as is seen in CS.
As mentioned in \S~\ref{subsubsec:iso-ratio} for the $^{13}$CS and C$^{34}$S emissions, the quality of the \ce{H2CS} data is also not high enough to enable an unambiguous distinction between centrally peaked or centrally depleted distributions for these weak lines, and despite appearances in the image plane the emission is consistent with azimuthally symmetric profiles. Higher signal-to-noise ratio and higher resolution data would help in disentangling potential differences in beam-filling factors for the different transitions that could impact the analysis.

\begin{figure}
\begin{center}
\includegraphics[scale=0.9]{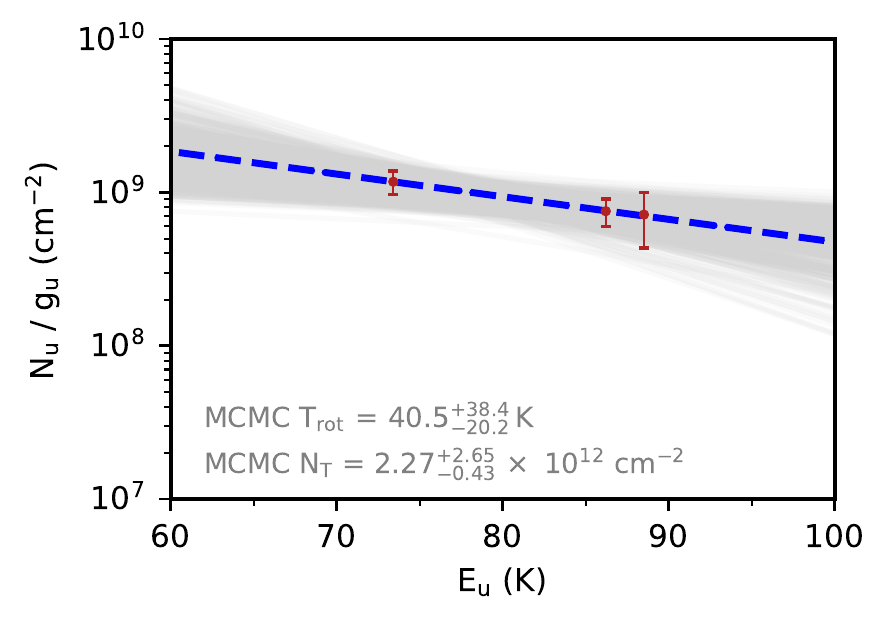}
\caption{Ortho-\ce{H2CS} rotational diagram built from the observed fluxes of the three ortho-\ce{H2CS} transitions $8_{17}-7_{16}$, $9_{19}-8_{18}$ and $9_{18}-8_{17}$ detected toward MWC~480, integrated over $\sim300$~au.
\label{fig:mwc480-H2CS-rot-diag}}
\end{center}
\end{figure}

\ce{H2CS} is an asymmetric rotor with two identical hydrogen nuclei. The molecule is thus found in two different nuclear spin configurations, {\it ortho} (where the quantum number $K_a$ is odd) and {\it para} (where the quantum number $K_a$ is even). The three \ce{H2CS} lines presented here are ortho lines, which can be used to derive a column density of the $o$-H$_2$CS, and a total column density assuming an $o/p$ ratio. This is done for MWC~480 in Fig.~\ref{fig:mwc480-H2CS-rot-diag} using the rotational diagram method described Sect.~\ref{subsubsec:pop-dia}. 
The following flat priors were assumed:
\begin{eqnarray}
    \tex (\rm{K}) = \mathcal{U}(10,100),\\
    N_{\rm {tot}} (\cc) = \mathcal{U}(\dix{7},\dix{20}).
\end{eqnarray}

Random draws and results from the posterior distributions are depicted in gray in Fig.~\ref{fig:mwc480-H2CS-rot-diag}, along with the least square fit results, shown in blue. The results converged toward 
\{$\tex\simeq41^{+38}_{-20}$~K, $N_{\rm {tot}}\simeq 3.0^{+2.7}_{-0.4}\tdix{12}\cc$\}, assuming the statistical ortho-to-para ratio of 3, toward the MWC~480 disk. The large uncertainties are due to a combination of low signal to noise and closely spaced energy levels. Still, it is noteworthy that the best-fit rotational temperature is close to the rotational temperature of CS, assuming CS is optically thin.

We used Eqs.~\ref{eq:Nu} and \ref{eq:boltzmann_dist} to derived an upper limit of $\ntot\lesssim2\tdix{12}\cc$ when considering the three \ce{H2CS} line tentatively detected toward LkCa~15, a rotational temperature of $T\approx 40$~K and an ortho-to-para ratio of 3. This corresponds to less than a factor of 2 lower than the value observed toward MWC~480.

\section{Astrochemical modeling}
\label{sec:modeling}

To explore the origin of the observed sulfur chemistry properties across our disk sample in general, and in MWC~480 and LkCa~15 in particular, we use a 1+1D protoplanetary disk model, based on the gas-grain chemical model \texttt{Nautilus (v.1.1)} \citep{wakelam2016}, tuned to the physical structures of MWC~480 and LkCa~15.

\subsection{Physical structure}

Assuming azimuthally symmetric disks, PPD physical parameters are commonly described in cylindrical coordinates centered on the inner star with a radius $r$ and the vertical axis $z$ perpendicular to the disks. Thus, in the following, we describe our PPD modeling along these two radial and vertical axes. The disk models extend from an inner radius of 1~au to an outer radius of 400~au. The other physical parameters used to compute the physical structures of MWC~480 and LkCa~15 are listed in Table~\ref{tab:physical-strcuture-disk-model}. We use the previously calculated temperature and density profiles from \cite{pietu2007} and \cite{guilloteau2011},assuming that the temperature and surface density, governing line emissions, vary as power laws of the radii \citep{beckwith1990,pietu2007}.
Below we describe the parameterization of temperature and density, visual extinction and UV fields, we used for our 1+1D disk modeling.

\subsubsection{Temperature profile}
\label{sec:ture-profile}

At a given radius $r$ from the central star, the vertical temperature gradient is computed following the modified prescription of \cite{dartois2003} used in \cite{rosenfeld2013} and \cite{williams2014}:

\begin{equation}
\small
T(z) = \left\{
    \begin{array}{ll}
        T_{\rm{mid}}+(T_{\rm{atm}}-T_{\rm{mid}})\left[ \sin \left(\frac{\pi z}{2z_q}\right) \right]^{2\delta}&\mbox{if} \, z<z_q\\
        T_{\rm{atm}}&\mbox{if} \, z\ge z_q,
    \end{array}
\right.
\label{eq:ture}
\end{equation}
where 
\begin{eqnarray}
T_{\rm{mid}}=T_{\mathrm{mid},R_c}\,\left(\frac{r}{R_{\rm{c}}}\right)^{-q},\\
T_{\rm{atm}}=T_{\mathrm{atm},R_c}\,\left(\frac{r}{R_{\rm{c}}}\right)^{-q},
\end{eqnarray}
vary as power law of the radii \citep{beckwith1990,pietu2007}. $R_c$ is a characteristic radius and $z_q=4H$, where $H$ is the pressure scale height modeled as the height at which the optical depth to UV photons, $\av$, becomes $\lesssim 1$. $H$, defined by the midplane temperature and assuming vertical static equilibrium, can be expressed as follows:

\begin{equation}
H=\sqrt \frac{k_{\rm{B}} \, T_{\rm{mid}} \,r^3}{\mu \,m_{\rm{H}}\, G \,M_\star},
\end{equation}
with $k_{\rm{B}}$ the Boltzmann constant, $\mu=2.4$ the reduced mass of the gas, $m_{\rm{H}}$ the proton mass, $G$ the gravitational constant, and $M_\star$ the mass of the central star. The midplane temperature $T_{\mathrm{mid},R_c}$ can be estimated from the following approximation for a simple irradiated passive flared disk:
\begin{equation}
    T_{\rm{mid}}(r)\approx \left(\frac{\varphi L_\star}{8\pi r^2 \sigma_{\rm{SB}}}\right)^{1/4},
    \label{eq:Tmid_Rc}
\end{equation}
with $L_\star$ the stellar luminosity, $\sigma_{\rm{SB}}$ the Stefan-Boltzman constant and $\varphi$ the flaring angle (e.g., \citep[e.g.][]{chiang1997,d'allesio1998,dullemond2001,dullemond2018,huang2018}.

The dust temperature is assumed equal to that of the gas. This is not a valid assumption in the uppermost layers of the disk \citep{bergin2007}, but is a reasonable simplification in this study focusing on S-chemistry in the 
molecular disk layers \citep[e.g.][]{woitke2009,akimkin2013, dutrey2017,semenov2018}. For instance, in the Flying Saucer edge-on T Tauri disk, the CS emission is observed to be less vertically extended than the CO one \citep{dutrey2017}. The resulting 2D temperature profiles for the MWC~480 and LkCa~15 disks are shown in the first column panels of Fig.~\ref{fig:Model-structure}.

\subsubsection{Density profile}
\label{sec:density-profile}

The disk is assumed to be in hydrostatic equilibrium.
 Thus, for a given vertical temperature profile (see sect.~\ref{sec:ture-profile}), the vertical density structure is determined by solving the equation of hydrostatic equilibrium:
\begin{equation}
\frac{\delta \ln \rho(z)}{\delta z}=-\left[\frac{G M_\star z}{r^3} \frac{\mu \, m_{\rm{H}}}{k_B T} + \frac{\delta \ln T}{\delta z}\right],
\label{eq:hidrostatic_eq}
\end{equation}
which gives for the isothermal case:
\begin{equation}
\frac{\delta \rho(z)}{\rho(z)}=-\left[\frac{G M_\star z}{r^3} \frac{\mu \, m_{\rm{H}}}{k_B T{_{\rm{mid}}}}\right]\delta z,
\end{equation}
with solution:
\begin{equation}
\rho(z)=\rho_o\exp\left(-\frac{z^2}{2H^2}\right),
\end{equation}
where $\rho_o$ corresponds to the volume density of the gas in the midplane. 
The surface density of the disk, $\Sigma$, can thus be expressed as follows:
\begin{equation}
\Sigma=\rho_0 \int_{-\infty}^{+\infty} \exp\left(-\frac{z^2}{2H^2}\right)=\sqrt{2\pi}\rho_0H.
\end{equation}

Furthermore, the surface density of the disk is assumed to follow a simple power law varying as $r^{-3/2}$ \citep{shakura1973,hersant2009} with sharp cutoff at specified inner and outer radii (i.e. $R_{\rm{int}}< r <R_{\rm{out}}$):
\begin{equation}
\Sigma (r) = \Sigma_c \left( \frac{r}{R_c}\right)^{-3/2},
\end{equation}
where $\Sigma_c$ is the surface density at the characteristic radius.

During its evolution, the disk loses materials toward the central star which can be quantified by the accretion rate $\dot{M} =-2\pi r \Sigma (r) v_r$, with $v_r$ the radial velocity of the disk. For axis-symmetric disk, mass conservation thus reads:
\begin{equation}
2\pi r \frac{\delta\Sigma}{\delta t}=-\frac{\dot{M}}{\delta r}, 
\end{equation}
If the mass loss of the disk is constant with time, i.e. $\frac{\delta \Sigma}{\delta t}=0$, the disk mass can simply be expressed as follows:
\begin{equation}
M_{\rm{disk}}=\int_{R_{\rm{int}}}^{R_{\rm{out}}} 2 \pi r \Sigma(r) dr.
\end{equation}

Thus, the surface density at a radius $R$ reads:
\begin{equation}
\Sigma_R=\frac{M_{\rm{disk}} R^{-3/2}}{4\pi \sqrt{R_{\rm{out}}}},
\end{equation}
which highlights the link between the mass of the disk, $M_{\rm{disk}}$, and its outer radius, $R_{\rm{out}}$.

The resulting 2D density profile is represented in the second column panels of Fig.~\ref{fig:Model-structure}, for each model.

\subsubsection{Visual extinction profile}
\label{sec:Av-profile}
The vertical visual extinction gradient is computed from the hydrostatic density profile using the conversion factor $N_{\ce{H}}/A_{\rm{V}}=1.6\tdix{21}$ \citep{wagenblast1989}, with $N_{\ce{H}}=N(\ce{H})+2N(\ce{H2})$ the vertical hydrogen column density of hydrogen nuclei. This conversion factor assumes a typical mean grain radius size of 0.1~$\mu$m and dust-to-mass ratio of 0.01, consistent with model assumptions. The resulting 2D visual extinction profile is represented in the third column panels of Fig.~\ref{fig:Model-structure}, for each model.

\subsubsection{UV flux distribution}

In the disk model the simplifying assumption of parameterizing the UV flux from the star as a multiple of the standard ISM radiation field in \cite{draine1978}'s units is done. This is a reasonable assumption for UV excess spectra of Herbig Ae stars, which are found to have UV field shapes similar to the interstellar radiation field (ISRF) \citep{chapillon2008}. For T Tauri stars UV excess spectrum can be dominated by accretion luminosity and therefore Lyman alpha, instead. Thus, for T Tauri stars, we are probably underestimating UV penetration into the disk, since Ly-alpha photons can scatter more efficiently into the disk \citep{bergin2003}. However, for simplicity and ease of comparison we use this approximation in both disk models presented here.

The unattenuated UV flux factor, $f_{\rm{UV}}$, at a given radius $r$ depends on both the photons coming directly from the central embedded star and on the photons that are downward-scattered by small grains in the upper atmosphere of the disk. 
Since the physical structure is built
from a parametric model approach (i.e. without radiative transfer treatment), the input UV flux of reference $f_{\rm{UV},R_c}$ (i.e. the UV flux factor at a radius $R_c$, see Table~\ref{tab:disk-params}), is divided by two \citep[e.g.][]{wakelam2016}. This 0th order approximation assumes that half of photons irradiating from the embedded young star diffuse into the disk, and half in the perpendicular direction. The UV flux factor at a given radius $r$ thus reads:
\begin{equation}
f_{\rm{UV}}=\frac{f_{\rm{UV},R_c}/2}{\left(\frac{r}{R_c}\right)^2+\left(\frac{4\rm{H}}{R_c}\right)^2}.    
\end{equation}

\begin{figure*}
\centering
\includegraphics[scale=0.42]{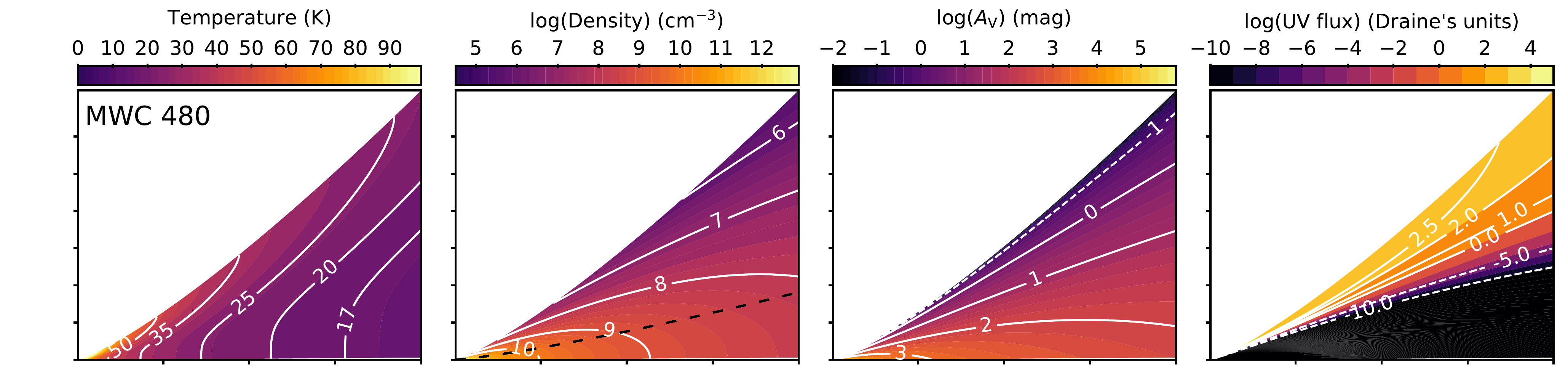}\\
\includegraphics[scale=0.42]{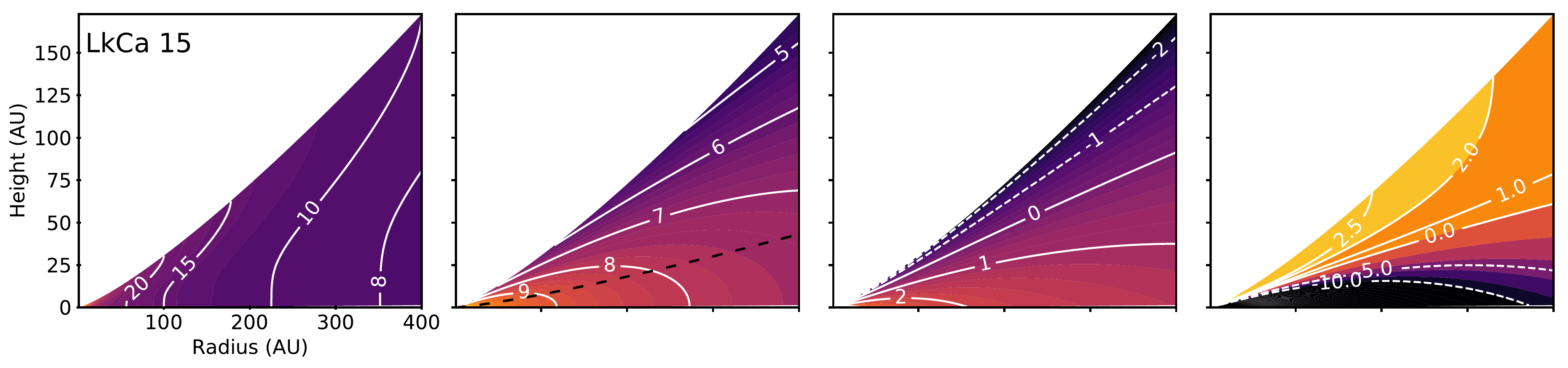}\\
{\caption{MWC~480 and LkCa~15 physical structures fed in our disk astrochemical model, based on the physical parameters listed in Table~\ref{tab:physical-strcuture-disk-model}. The 2D temperature ({\it first column}), density ({\it second column}), visual extinction ({\it third column}) and UV flux ({\it fourth column}) profiles are represented as functions of disk radius versus height, both in au. The dashed black line, on the densities panels, delineates 1 scale height.}
\label{fig:Model-structure}}
\end{figure*}

To calculate the UV flux at any point in the disk the UV flux factor is convolved with the visual extinction profile. The resulting 2D UV flux is represented in the last column panels of Fig.~\ref{fig:Model-structure}, for each model.

\begin{table*}
\begin{center}
\caption{Physical parameters used for our disk models  \label{tab:physical-strcuture-disk-model}}
\begin{tabular}{lcc}
\hline\hline
Parameters&MWC~480$^a$&LkCa~15$^b$\\
\hline
\hline
Stellar mass: $M_\star$ ($M_\odot$) &1.8&1.0\\
Disk mass: $M_{\rm{d}}$ ($M_\odot$)& 0.18&0.03\\
Characteristic radius: $R_{\rm{c}}$ (AU) &100&100\\
Outer cut-off radius: $R_{\rm{out}}$ (AU) &500&550\\
Density power-law index: $\gamma$&1.5&1.5\\
Midplane temperature at $R_{\rm{c}}$$^c$: $T_{\rm{100AU}}$(K) &30&15\\
Atmosphere temperature at $R_{\rm{c}}$: $T_{\rm{100AU}}$(K) &48&20\\
Surface density at $R_{\rm{c}}$: &5.7&0.9\\
Temperature power-law index: $q$ &0.5&0.5\\
Vertical temperature gradient index: $\beta$&2&2\\
UV Flux of reference: $f_{{\rm UV},R_c}$ (in \cite{draine1978}'s units) &8500$^d$&2550$^e$\\
\hline
\hline
\end{tabular}
\begin{list}{}{}
\item $^a$ \cite{guilloteau2011}
\item $^b$ \cite{pietu2007}
\item $^c$ The midplane temperatures are estimated from Eq.(~\ref{eq:Tmid_Rc}), the luminosities listed Table~\ref{tab:disk-params} and a typical flaring angle $\varphi=0.05$.
\item $^d$ from \cite{dutrey2011}, originally computed from the \cite{kurucz1993} ATLAS9 of stellar spectra .
\item $^e$ from \cite{dutrey2011}, originally scaled from \cite{bergin2004}. 
\end{list}
\end{center}
\end{table*}

\subsection{Chemical modeling}
\label{subsec:chemical_modeling}

Once built, the 
1+1D physical structure described above (and depicted in Fig.~\ref{fig:Model-structure} for the MWC~480 and LkCa~15 cases) is fed as an input in the pseudo 
time-dependent astrochemical modeling code \texttt{Nautilus v1.1} \citep{wakelam2016} used in three-phase mode \citep{ruaud2016}. This rate-equation gas-grain code -- based on \cite{hasegawa1992} and \cite{hasegawa1993} -- simulates the time-dependent chemistry of $\sim 1100$ species (half in the gas phase and half in solid phase) linked together via more than $\sim 12000$ reactions, in the vertical direction at each radius in three different phases: gas, grain surface (top two ice layers on grains), and grain mantle (deeper ice layers on grains). Exchanges in between all the different phases are included: adsorption and desorption processes link the gas and surface phases, and swapping processes link the mantle and surface of grains. Several desorption mechanisms are considered, thermal desorption \citep{hasegawa1992}, and non thermal ones, such as cosmic-ray induced desorption \citep{hasegawa1993}, photodesorption and chemical desorption \citep[for further details see e.g.][]{garrod2007,ruaud2016,wakelam2016,legal2017}. For photodesorption we use the prescriptions discussed in \cite{legal2017}, where we followed the recommendations of \cite{bertin2013} to use a simplistic approach. This consists in considering a single photo-desorption yield for all the molecules rather than individual ones, only experimentally determined for a few number of species. Thus, we set a constant generic photo-desorption yield of $1\tdix{-4}$ molecule/photon \citep{andersson2008} for all the species contained in our chemical network.
In the gas phase typical bi-molecular ion-neutral and neutral-neutral reactions are considered, as well as cosmic-ray induced processes, photoionizations and photodissociations caused by both stellar and interstellar UV photons. 
A dust-to-mass ratio of 0.01 is assumed, with spherical grains of $0.1~\mu$m single radius size. 

A key parameter of the new modeling study we present here, is the use of an up-to-date sulfur chemical network, based on the KInetic Database for Astrochemistry (KIDA)
(http://kida.obs.u-bordeaux1.fr/), and including recent updates \citep{vidal2017,legal2017,fuente2017}, that are relevant for the purpose of this work.


\begin{table}
\centering
\scriptsize
\caption{Initial Elemental Abundances}
\begin{tabular}{lcc}
\hline\hline
Species & $n_i/n_{\text{H}}$    & Reference\\
\hline
H$_2$   & 0.5\\
He      & 9.0$\times$10$^{-2}$              & 1\\
C$^+$   & 1.7$\times$10$^{-4}$              & 2 \\
N       & 6.2$\times$10$^{-5}$              & 2\\  
O       & 2.4$\times$10$^{-4}$              & 3\\ 
S$^+$   & 8.0$\times$10$^{-8}$   & 4 \\
Si$^+$  & 8.0$\times$10$^{-9}$              & 4 \\
Fe$^+$  & 3.0$\times$10$^{-9}$              & 4 \\
Na$^+$  & 2.0$\times$10$^{-9}$              & 4 \\
Mg$^+$  & 7.0$\times$10$^{-9}$              & 4 \\
P$^+$   & 2.0$\times$10$^{-10}$             & 4 \\
Cl$^+$  & 1.0$\times$10$^{-9}$              & 4 \\
F$^+$   & 6.7$\times$10$^{-9}$             & 5 \\
\hline
\label{tab:elemental_ab}
\end{tabular}
\tablecomments{
(1) \cite{wakelam2008}; (2) \cite{jenkins2009};
(3) \cite{hincelin2011};
(4) \cite{graedel1982};
(5) \cite{neufeld2015};}
\end{table}

To take into account 
chemical inheritance from previous 
stages, we first simulate the chemical evolution of a starless dense molecular cloud up to a characteristic age of $1\tdix{6}$ years \citep[e.g.][]{emelgreen2000,hartmann2001}. 
For this 0D model typical constant physical conditions were used: grain and gas temperatures of 10~K, a gas density $n_{\rm{H}}=n(\ce{H})+n(\ce{H2})=2\tdix{4} \ccc$ and a cosmic ionization rate of $1.3\tdix{-17}\s$; this parent molecular cloud is also considered to be shielded from external UV photons by a visual extinction of 30 mag. For this first simulation stage, we consider diffuse gas starting conditions, i.e. that initially  all  the  elements  are  in  atomic  form (see  Table~\ref{tab:elemental_ab})  except  for  hydrogen assumed to be initially already fully molecular. The elements taken into account in our simulation with an ionization potential lower than that of hydrogen (13.6 eV) are thus assumed to be initially singly ionized, see Table~\ref{tab:elemental_ab}. The chemical gas and ice compositions of this representative parent molecular cloud serve as the initial chemistry for our 1+1D disk model.

We ran the disk chemistry models for one million years. This is on the low side for our disk sample (see Table~\ref{tab:disk-params}), but has been shown in previous studies to be a reasonable approximation of the chemical age of a disk when grain growth is not taken into account \citep[e.g.][]{cleeves2015}. Even though the chemistry in the disk has not reached steady state at such a time, its evolution is slow enough that the results presented here hold for a disk twice younger or older. We note that the purpose of these models is not to quantitatively fit observations, but rather to obtain a first intuition on whether existing sulfur chemistry networks can reproduce observed sulfur species.

 With the disk model presented above, we thus ran two different disk chemistry simulations that differentiate from one another only by their physical structures: one simulating the MWC~480 disk and a second the LkCa~15 disk (see Fig.~\ref{fig:Model-structure}). Both models start from the same initial chemical composition, that results from the 0D model described above and considers a "low" sulfur elemental abundance of $8\tdix{-8}$ \citep[see Table~\ref{tab:elemental_ab},][]{graedel1982,wakelam2008},
 i.e. $\sim2$ orders of magnitude below the cosmic value $\sim1.5\tdix{-5}$ \citep{asplund2009}. This "low" sulfur elemental
 abundance corresponds to the typical sulfur depleted value 
used in models to reproduce the abundances observed in dense astrophysical molecular environments \citep[e.g.][]{tieftrunk1994,wakelam2004,vidal2017,vastel2018}. For the present study, this approximation seems reasonable since CS and \ce{H2CS} are supposed to reside in disk molecular layers, which are similar to dense astrophysical molecular environments.


\begin{figure*}
\begin{center}
\includegraphics[scale=0.38]{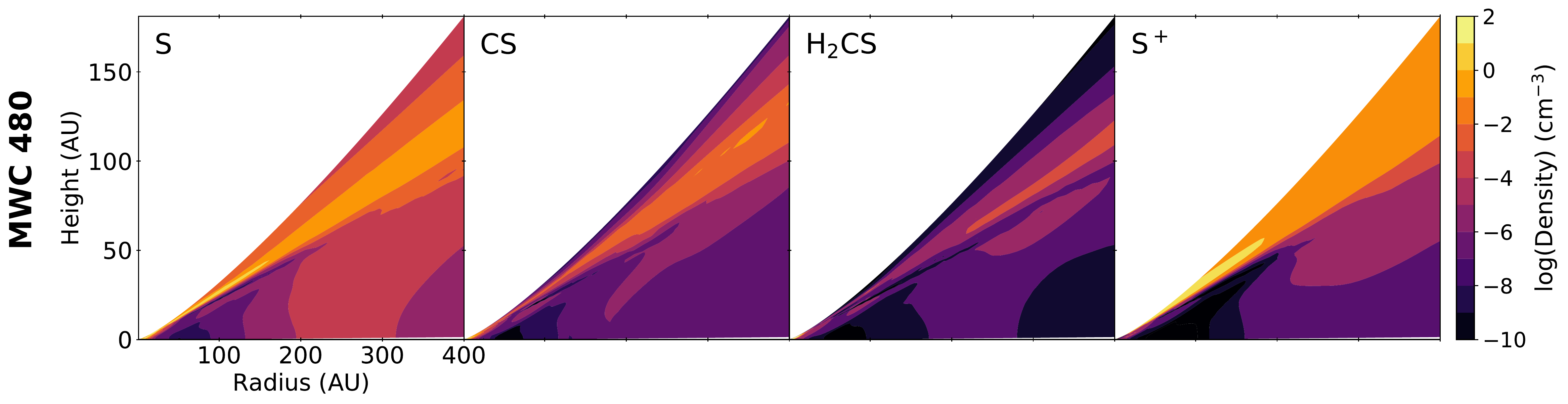}
\includegraphics[scale=0.38]{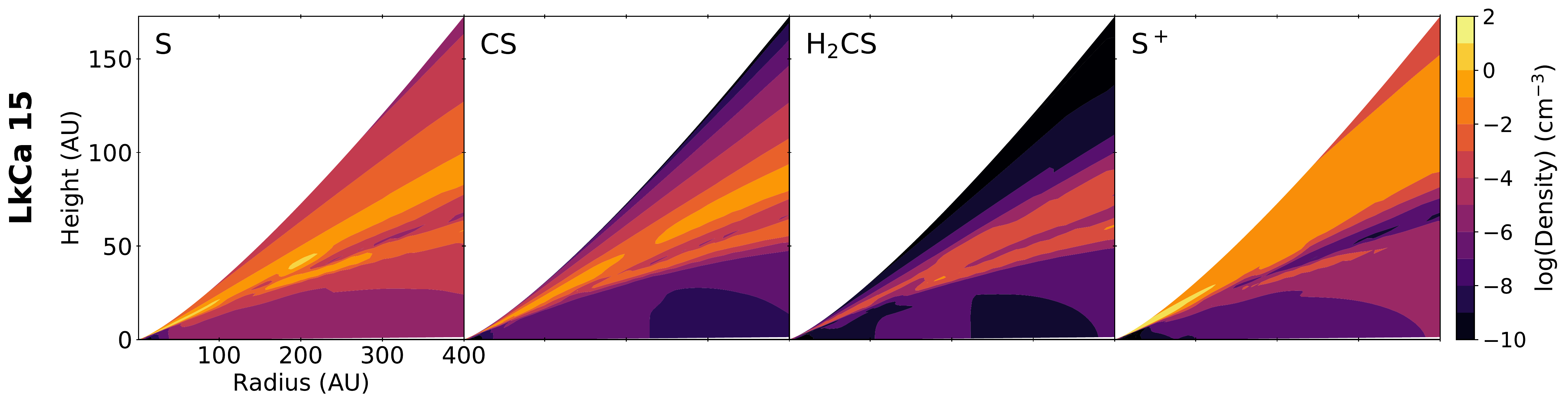}
\caption{Modeled height versus radius number densities (i.e. absolute abundances) of CS, \ce{H2CS}, \ce{S+}, and S. {\it Top panel:} MWC~480 model, {\it Bottom panels:} LkCa~15 model. \label{fig:2D-modelled-dens}}
\end{center}
\end{figure*}
\subsection{Model results}
\label{subsec:model-results}



The resulting 
number densities of CS, \ce{H2CS} and their main precursors, \ce{S+} and S, 
at 1~Myr 
are presented in Fig.~\ref{fig:2D-modelled-dens} for the MWC 480 and LkCa 15 disk models (the corresponding fractional abundances
are shown in Appendix~\ref{app:modelled-fractional-ab}). 
Fig.~\ref{fig:2D-modelled-dens} shows that
in the warmer and denser Herbig Ae disk MWC~480, all gas-phase S carriers occur in higher disk layers than in the colder T Tauri disk LkCa~15 (see Table~\ref{tab:physical-strcuture-disk-model}). Even if a common expectation is that the higher UV fluxes found around more massive stars will push the molecular layers down to deeper disk layers, since the disk is more massive,  the visual extinction is much larger (as can be seen in Fig.~\ref{fig:Model-structure}), and so the overall outcome is that UV does not penetrate. MWC~480 has a factor of 6 larger disk mass than LkCa~15. In the more tenuous LkCa~15 disk the shielded layers thus occur deeper into the disk than would be expected if it was as massive as MWC 480.
Furthermore, chemistry is not just set by UV radiation, but also by density, which, e.g., controls the recombination rate. When comparing Fig.~\ref{fig:2D-modelled-dens} with Fig.~\ref{fig:Model-structure}, we see that the layers where CS and H$_2$CS reside have similar densities of $\sim \dix{6}-\dix{8}\ccc$.

In both disks the S and \ce{H2CS} layers reside deeper toward the disk midplane than \ce{S+} and \ce{CS}. The depth of the S/\ce{S+} transition is set by a balance of, on the one hand, production of S$^+$ from S through photo-ionization and charge transfer reaction with \ce{C+} and, on the other hand, recombination of S$^+$ with electrons to form S. The main formation pathway to form CS starts with S$^+$, while H$_2$CS mainly forms from atomic S, explaining their difference in vertical distribution.

In more detail, the CS-producing sulfur chemistry in the upper disk layers is driven by rapid ion-neutral reactions between \ce{S+} and small hydrocarbons \ce{CH$_x$} and \ce{C$_y$H} (with $x=1-4$ and $y=2-3$). This produces carbonated S-ions including \ce{HCS+}, \ce{CS+}, \ce{HC3S+}, and \ce{C2S+}, that subsequently recombine with electrons to form neutral S-containing species. CS is thus a daughter molecule of \ce{S+}. CS also has slower neutral-neutral formation pathways starting with S and neutral small hydrocarbons, which produces a second CS reservoir at deeper disk layers. Its main destruction pathways are photodissociation processes mainly in the disk atmosphere, reactions with protons and protonated ions (i.e. with \ce{H+}, \ce{H3+}, \ce{HCO+}) and with \ce{He+}, and freeze-out onto grain surfaces.

\ce{H2CS} is mainly produced by the gas-phase neutral-neutral reaction of atomic S and \ce{CH3} and the electronic dissociative recombination of \ce{H3CS+}, whose formation begins with the reaction between \ce{S+} and \ce{CH4}. At high densities and low temperatures, i.e. in regions closer to the midplane, some grain-surface reactions can also contribute to its formation, such as the hydrogenation reaction \ce{s-H + s-HCS -> s-H2CS} releasing some ($\sim 1\%$) of its \ce{H2CS} product in the gas-phase by chemical reactive desorption. Similarly to CS, \ce{H2CS} is mainly destroyed by photodissociation processes in the disk atmosphere, reactions with protons and protonated ions (i.e. with \ce{H+}, \ce{H3+}, \ce{HCO+}), and freeze-out onto grain surfaces.
\begin{figure*}
\includegraphics[scale=0.67]{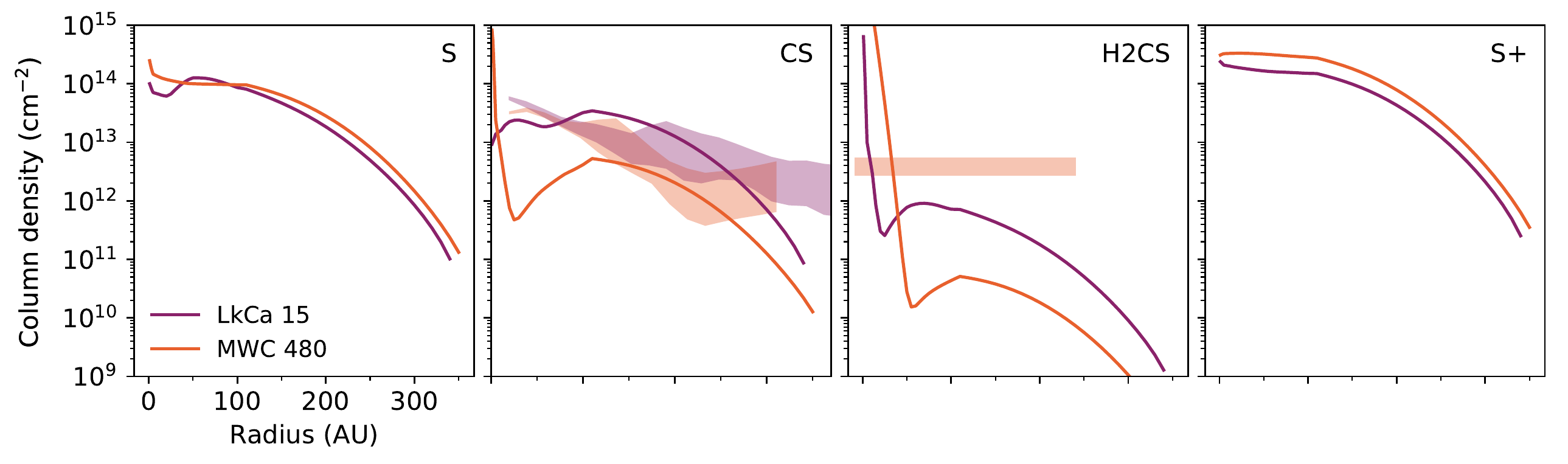}
\caption{S, CS, \ce{H2CS} and \ce{S^+} modeled column densities vertically integrated from the disk upper layer to the midplane and convolved to a resolution of $0.5''$ to facilitate the comparison with the observations. The modeled column densities are shown by the solid lines in dark purple for LkCa~15 and in orange for MWC~480. Observational error bars from the present study are represented by the transparent boxes in purple for LkCa~15 and in orange for MWC~480 (radially resolved for CS and disk-averaged for \ce{H2CS}).
\label{fig:coldens-plot-radius}}
\end{figure*}

\begin{figure}
\includegraphics[scale=0.9]{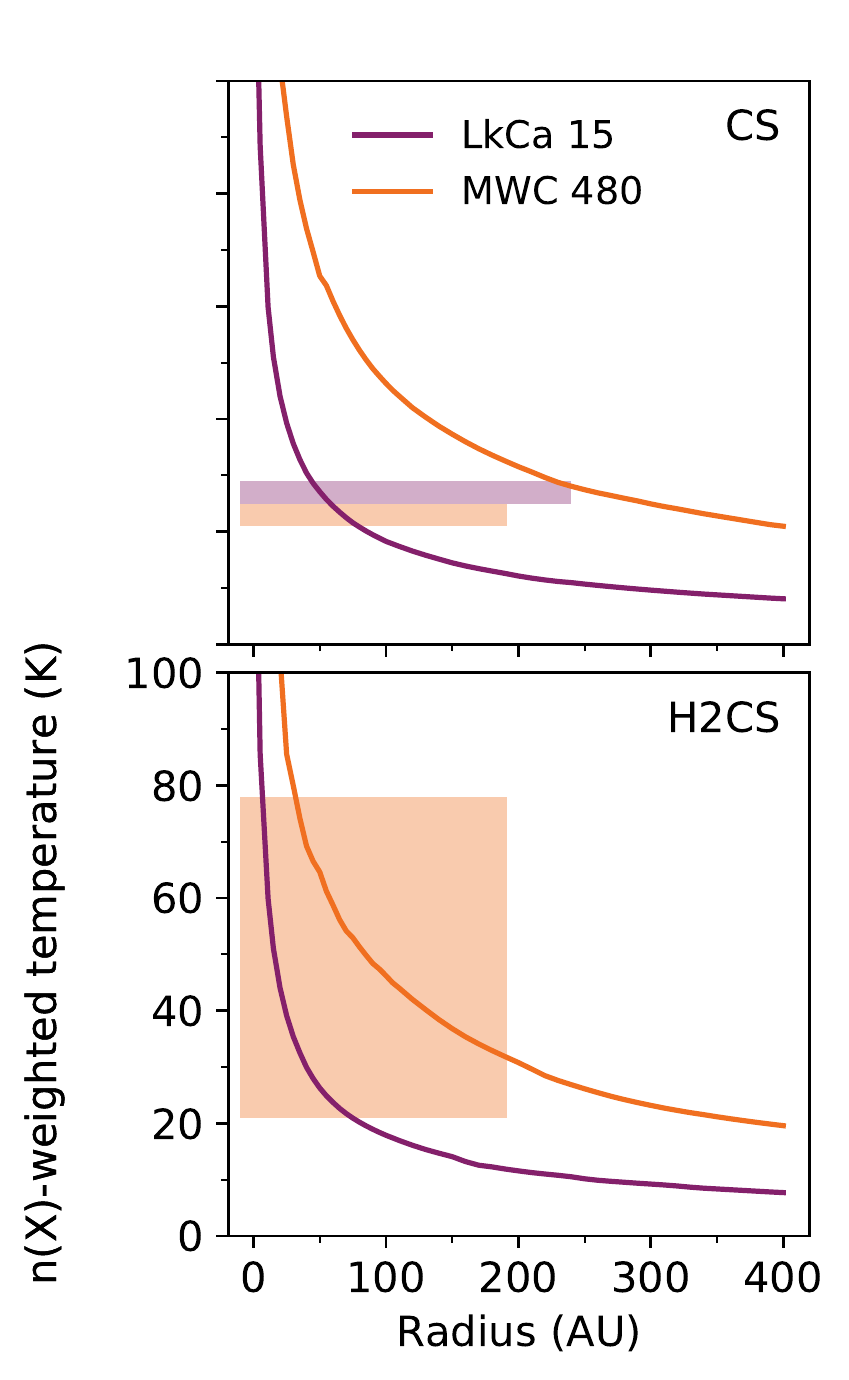}
\caption{Modeled mean disk gas temperature weighted by the density of S, CS, \ce{H2CS} and \ce{S^+} - as a function of the radius from the central disk object. The models are represented by the solid lines in dark purple for LkCa 15 and in orange for MWC 480. Observational error bars from the present study are overplotted in horizontal transparent boxes in purple for LkCa 15 and in orange for MWC 480. \label{fig:spec-dens-weigted-ture}}
\end{figure}


The resulting radial distributions of the vertically integrated column densities along with the observationally constrained column densities are presented in Fig.~\ref{fig:coldens-plot-radius}. 
For both the  MWC~480 and LkCa~15 models, the CS column density peaks at a radius of $\sim100$~au, and decreases from there with increasing radius. The theoretical outer disk CS column density profiles appear similar for both disks, but in the inner disk, the LkCa~15 column density dips, while the MWC~480 column density peaks. Apart from this inner peak, the theoretical LkCa~15 column densities are systematically higher than those of MWC~480, by a factor of $\sim2-10$, consistent with observations. 
Across the disks, the model predictions are generally within a factor of a few for both MWC~480 and LkCa~15, which must be considered as good agreement, considering that there is no tuning of the model involved.
We also note, that the CS column density derived from the  $^{13}$CS observations toward LkCa~15 being an order of magnitude higher are thus inconsistent with our model predictions. However, we note that our disk models were run with an initially depleted S-element abundance (see \S~\ref{subsec:chemical_modeling}). Running the same models with undepleted S-element abundance results in an overprediction of CS by 1-2 orders of magnitude, while \ce{H2CS} is in better agreement compared to the depleted models. This emphasizes the need for additional S-species disk observations to anchor S-disk chemistry modeling.

The predicted H$_2$CS column densities are 1--2 orders of magnitude below the predicted CS column densities in each disks.
Similarly to CS, the H$_2$CS profiles are characterized by fall-off in the outer disk, a local maximum at $50-100$~au, and then (similar to CS in MWC~480) an increasing column density toward the central source. The dips in the inner disks are not seen in the CS and H$_2$CS precursors, with the possible exception of S toward LkCa~15. 
Compared with observations toward MWC~480, the model underpredicts H$_2$CS by several orders of magnitude. Furthermore the model predicts that more H$_2$CS should be present toward LkCa~15 compared to MWC~480, which is opposite to what we observed. Finally the model predicts that H$_2$CS should be centrally very peaked, while observations suggest more extended emission.


Figure~\ref{fig:spec-dens-weigted-ture} presents the calculated mean disk gas temperature weighted by the density of CS and H$_2$CS, respectively in the two disks, which provides a measure of the typical temperature environment within which these molecules reside.  
In the outer disk (i.e. for radius $> 200$ au), CS resides at temperatures in the range $\sim 8-12$~K and $\sim 20-30$~K, in the LkCa~15 and MWC~480 disk models, respectively. In the inner disk (i.e. for radius $\lesssim 50$ au) they reside at temperatures in the ranges $\sim25-200$~K and $\sim60-400$~K, respectively.
Similar behaviors can be seen for \ce{H2CS}. While the H$_2$CS density 
peaks in more embedded disk layers, the vertical temperature gradient is small in our model, and the difference in temperature when compared with CS is therefore also small (see Fig.~\ref{fig:2D-modelled-dens}). The rotational temperatures of CS and \ce{H2CS} are also overplotted on Fig.~\ref{fig:spec-dens-weigted-ture}. These temperatures were derived from observations out to $\sim$190~au (MWC~480) amd $\sim$240~au (LkCa~15), and were found to be quite similar. Compared to CS observations, the models overpredict the temperature for MWC~480, and underpredict it for LkCa~15.


\section{Discussion}
\label{sec:discussion}



\subsection{Model predictions vs observations}
\label{subsec:model-vs-obs}

\begin{figure*}
\begin{center}
\includegraphics[scale=0.38]{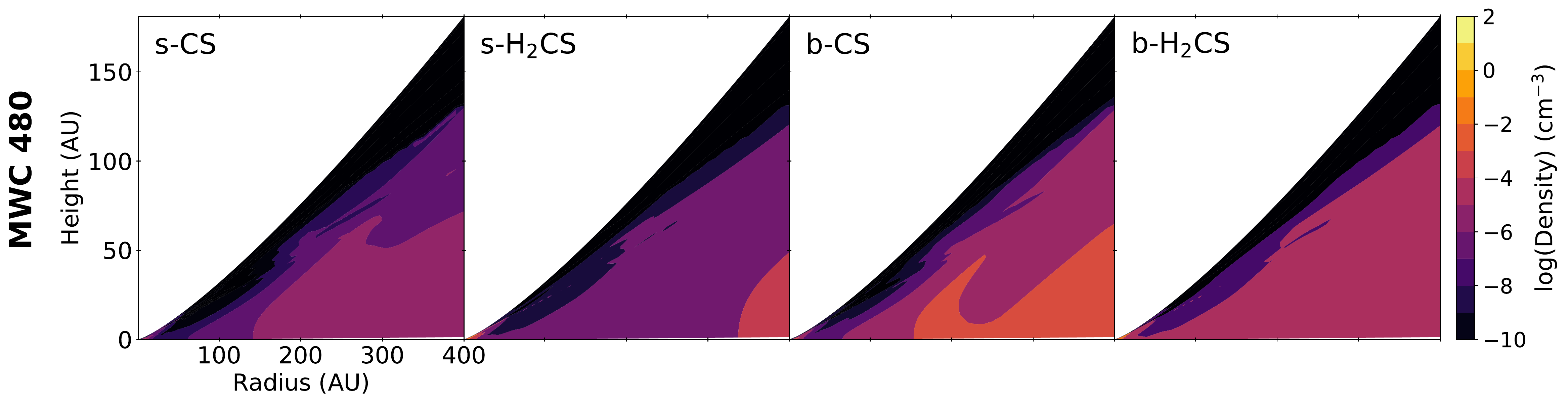}
\includegraphics[scale=0.38]{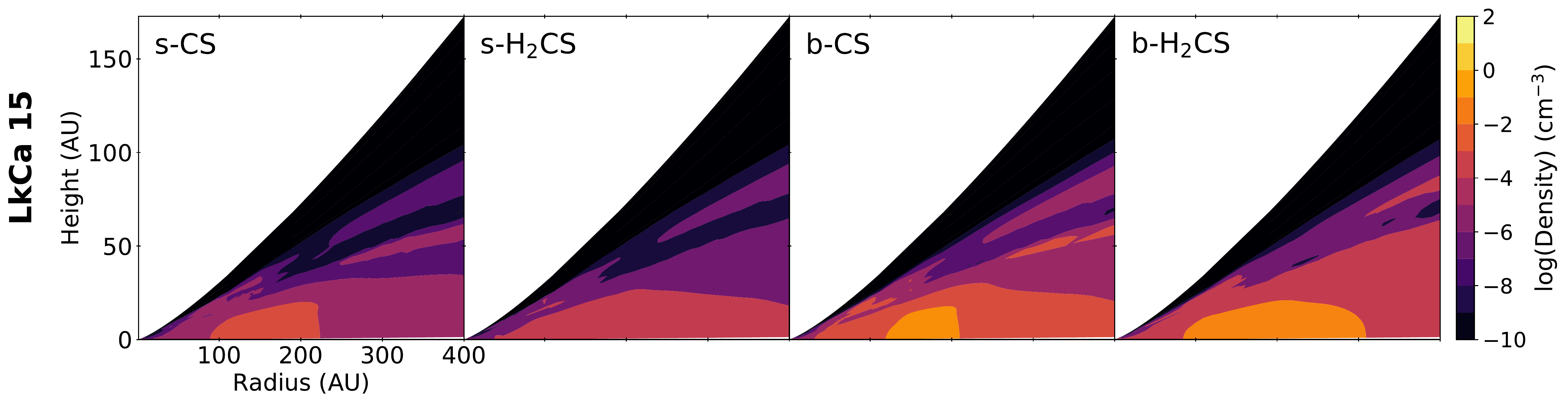}
\caption{Modeled height versus radius number densities (i.e. absolute abundances) of surface (s-) and bulk (b-) CS and \ce{H2CS}. {\it Top panel:} MWC~480 model, {\it Bottom panels:} LkCa~15 model. \label{fig:2D-modelled-dens-dust}}
\end{center}
\end{figure*}

In \S~\ref{subsec:model-results} we compared model predictions of CS and H$_2$CS column densities, radial profiles and excitation temperatures. In this section we summarize agreements and disagreements and discuss possible explanations for cases where there is a mismatch between observations and theory.

{\it Column densities:} The observed CS column densities of 10$^{12}$--10$^{13}\cc$ are well reproduced, within a factor of a few, by our models. The best fit are found for radii in the range $\sim150-250$~au for MWC~480, and for radii of $\sim100$ au and in the range $\sim200-275$~au for LkCa~15. However, we should note the discrepancies obtained for the inner disk of MWC~480 (i.e. radii~$\lesssim100$~au) between the CS column density profiles derived from the observations and the one obtained from our model. Both higher angular resolution observations and further modeling investigations are required to disentangle this issue. By comparison, the models underpredict H$_2$CS in the outer disk of MWC~480, and incorrectly predict that H$_2$CS should be more abundant in LkCa~15 than in MWC~480. An explanation could be related to the fact that, for simplicity, we do not include LkCa15's wide inner cavity \citep[e.g.][and reference therein]{alencar2018} in the LkCa disk structure. This will likely impact some aspects of the chemistry model predictions, especially at radii of $<50$~au. In the outer disk, the model predictions should be more reliable, or at least not limited by this particular omission. Future disk modeling that takes into account this structural feature and its impact on radiative transport is needed to make quantitative model-data comparisons in the inner disk. A possible reason for the overall underprediction of H$_2$CS is that grain surface formation of H$_2$CS and subsequent photodesorption may be more important in disks than is currently implied by the model. The surface and bulk ice H$_2$CS 2D disk densities are represented in Fig.~\ref{fig:2D-modelled-dens-dust}, as well as those of CS, for both MWC~480 and LkCa~15 disk model. We see that both molecules are abundant in ices, especially in the disk midplanes. However, photo-irradiation laboratory experiments of \ce{CH3OH} ices showed an efficient dissociation of the product rather than a desorption \citep{bertin2016,cruz-diaz2016}. \ce{H2CS} being also a rather large molecule, it could similarly efficiently be photo-dissociated in the ice. Our present model already considers ice photo-dissociation rates, however, due to the lack of laboratory studies on that topic, those are set to be equal to the gas-phase photo-dissociation ones. So, our model might misspredict the \ce{H2CS} ice abundance and desorption, unless additional ice formation pathways are missing in our astrochemical networks. Therefore, both, photo-irradiation of icy \ce{H2CS} and new \ce{H2CS} ice formation pathways require laboratory experiments.
Moreover, the presented model does not take into account grain growth and migration, which may have depleted the outer disk of UV protection, increasing both its temperature \citep{cleeves2016} and the importance of ice photodesorption in this region \citep{oberg2015}. This needs to be revisited in a future model that incorporates this structural element, where the ISRF impact may increase if the outer disk is more tenuous. Interestingly, the MWC~480 continuum disk appears more compact than the LkCa~15 disk, and ice photodesorption may therefore be important in a larger portion of the MWC~480 disk, possibly explaining why it hosts more H$_2$CS in the gas-phase compared to the LkCa~15 disk. In addition, the sulfur chemistry network is probably still incomplete, and there may be other gas- and grain-surface pathways to \ce{H2CS} that are not yet considered in the model. Finally, we should also note that beam dilution could be an additional explanation for the mismatch observed in between the \ce{H2CS} observations and its predictions from the astrochemical disk modeling. Thus, higher \ce{H2CS} angular resolution observations would be required toward the same disk sample to further investigate these issues.


{\it Radial profiles:} Our beam size corresponds to 80~au, and we are therefore not sensitive to radial scales below 40~au. On larger scales there is qualitative agreement between observed and modeled CS radial profiles, including the central dip seen toward MWC~480. However, the modeled predicted dip appears deeper than the one derived from the observations when using the rotational diagram method (Fig.~\ref{fig:CS-radial-rot-dia}), but this might in part be a question of radiative transfer. However, in all other disks we observed a central CS peak. It would thus be interesting to further investigate if this is just a spatial resolution issue or if this characterizes a specific difference in between the Herbig Ae MWC~480 disk and the other T Tauri disks of our sample. To address this issue additional observations would be required, both in the same disk sample with higher angular resolution and toward additional Herbig Ae disks to increase our statistics on the latter type of disk. Another interesting feature to further explore is the break observed in CS at the continuum edges across our whole disk sample. To address this issue, the coupling between dust and gas need to be refined in our current modeling.

{\it Excitation temperatures:} Model predictions for CS excitation temperatures are too high for MWC~480 and too low for LkCa~15. This is probably a result of poorly constrained vertical temperature structures in disk models rather than an issue with chemistry model predictions under which conditions CS is present in the gas-phase; the CS chemistry is relatively straight forward and should mainly depend on the presence of ionized S, and gas-phase carbon. We suspect that the LkCa~15 disk is modeled to be too cold. The model temperature structures of both disks were calculated using some rather simplistic model assumptions and need to be updated to fit more recent observations than the ones available. This will be the subject of a future theory-focused study. According to the model predictions, the S-bearing species presented in Fig.~\ref{fig:2D-modelled-dens} reside at intermediate scale heights for the LkCa~15 disk and at slightly higher ones for the warmer disk MWC 480. This could result from the too simplistic physical structure modeling performed in the present study and would benefit from physical structure modeling refinement and improvement. However, as we mentioned in \S~\ref{subsec:model-results}, the MWC~480 disk being more massive, its visual extinction is larger (see Fig.~\ref{fig:Model-structure}), implying less UV penetration in the disk. In addition, the dramatic changes in temperature over small radii shown in Fig.\ref{fig:spec-dens-weigted-ture} highlight, once again, that our observations suffer from insufficient spatial resolution to effectively trace this trend, and therefore higher resolution CS observations are required to test this model result.

High spatial resolution observations of additional sulfur-bearing molecules in the same disks  are needed to better constrain disk sulfur chemistry. Furthermore, observations toward other Herbig Ae disks would also help in improving our sulfur chemistry knowledge since, according to the comparison between the MWC~480 observations presented here versus the T Tauri ones, there might be some sulfur chemistry distinctions in between T Tauri and Herbig Ae disks.

Also, we should note that the impact of X-rays on disk chemistry is not included in our present model. X-rays could affect the 
the ion abundances \citep[e.g.][]{glassgold1997,aikawa1999,henning2010,teague2015,cleeves2015} such as the abundances of \ce{S+}, \ce{CS+}, \ce{HCS+} as well as the SO/\ce{SO2} formation, and thus affects the overall S-chemistry \citep[e.g.][]{semenov2018}. However, for the ALMA observations presented here, concerning disk radii beyond $\sim10-20$ ~au, X-rays chemistry is not dominant \citep[e.g.][]{rab2018}. Nevertheless, this would be an interesting future modeling improvement.
\subsection{S-carriers in disks and other astrophysical environments}


Compared to other disk studies, the CS column densities derived toward our disk sample (see Fig~\ref{fig:CS_coldens_full_sample}) are consistent with the values found in other disks, e.g. $2.0\pm0.16\tdix{12}\cc$ in GO Tau and $8.7\pm1.6\tdix{12}\cc$ in LkCa~15 \citep{dutrey2011},
$6\pm3\tdix{12}\cc$ in DM Tau \citep{semenov2018} and $2.2\tdix{13}\cc$ in GG Tau \citep{phuong2018}, which agrees with CS being the most abundant S-species detected in disks so far.

\begin{figure*}
\begin{center}
\includegraphics[scale=0.45]{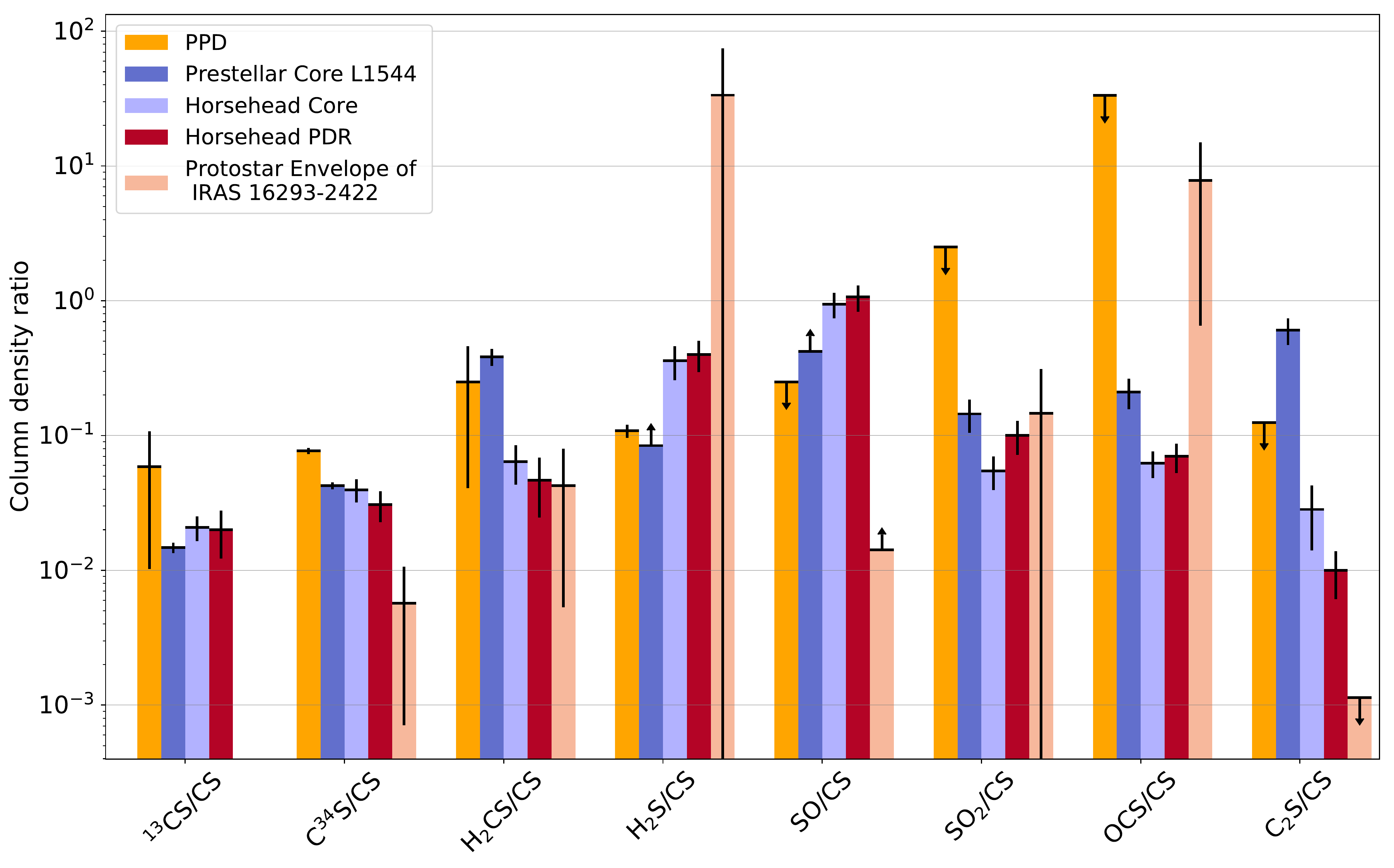}
{\caption{Column density ratios of the S-species observed in protoplanetary disks (PPD) (\cite{chapillon2012,phuong2018,semenov2018}, and this work) versus the gas S-reservoir in disk CS. These ratios are compared to the values observed in the prestellar core L1544 \citep{vastel2018}, in the IRAS-16293-2422 protostar envelope \citep{drozdovskaya2018}, and in the Horsehead nebula (Riviere-Marichalar et al. submitted). 
The error bars are represented by the black crosses and the upper and lower limits by the black arrows. \label{fig:histo}}}
\end{center}
\end{figure*}

In order to place our CS and \ce{H2CS} disk observations in a more hollistic view of sulfur chemistry, we explore in this section what characterizes the sulfur chemistry in different interstellar environments, based on observations from the literature reported in PDR (Riviere-Marichalar et al. submitted), molecular clouds \citep{drozdovskaya2018} and cold dense core \citep{vastel2018}, analogs of the main three chemical layers constituting protoplanetary disks, i.e. the disk atmosphere, intermediate molecular layers, and the disk midplane. Figure~\ref{fig:histo} presents the column density ratios of the main S-bearing species observed thus far in PPD (from this work and previous studies) versus CS, compared to their values in the different ISM environments aforementioned.

The $^{13}$CS/$^{12}$CS value we derived in disks from this study is found higher ($\sim 0.06$) than those measured so far in the different ISM environments presented in Fig.~\ref{fig:histo} ($\sim 0.01-0.03$). However, the wide error bars we derived ($\pm 0.05$) do not allow us to draw strong conclusions on the plausible $^{13}$C enhancement suggested in this work. Higher resolution observations of CS isotopolgues in disks are needed to further investigate this $^{13}$C enhancement hint.

The C$^{34}$S/C$^{32}$S ratio we observed in disks ($\sim 0.08$) is found, within error bars, a factor of 2 higher than values found in PDRs and dense cores, and than the LISM value. The observed C$^{34}$S/C$^{32}$S ratio found in this work is also about an order of magnitude higher compared to the value found in the molecular envelope of the IRAS~16293-2422 Sunlike protostar. Further observational explorations of this ratio from protostar envelopes to disks could thus bring some clues on the S-chemistry, both, on the question of chemical inheritance versus chemical reset during the stellar formation and on the physical environment impact (UV flux, temperature, etc.) on it.

The \ce{H2CS}/CS ratio decreasing from dense to diffuse ISM, is found in disks (i.e. in LkCa~15) almost as high as in dense cores \citep{vastel2018}. Nevertheless, the high uncertainties on our measurement do not allow us to firmly assert that the \ce{H2CS}/CS ratio behaves in disk more like dense core S-chemistry than diffuse ISM S-chemistry. To arbitrate on that, higher resolution observations are also needed. 

Interestingly, and in opposition to the \ce{H2CS}/CS ratio, the \ce{H2S}/CS is increasing from dense to diffuse ISM and the disk value \citep{phuong2018} is found almost as low as the lower limit derived in dense core \citep{vastel2018}. This could also explain why \ce{H2S} was so hard to detect in disks \citep[e.g.][]{dutrey2011}, aside from the fact that the \ce{H2S} detection could have been facilitate by the high mass of the GG Tau disk in which it was detected. Also, even though high uncertainties remain on value observed toward IRAS 16293-2422, \ce{H2S} appears as one of the main S-reservoirs in this source. A similar result is found for the OCS/CS ratio toward IRAS 16293-2422. These last two results could reflect more efficient thermal desorption in this source, that would also explain  the boost in S-chemistry observed there \citep{drozdovskaya2018}. However, and in opposition to the \ce{H2S/CS} ratio, the OCS/CS ratio is found lower in PDR than in dense core, suggesting that the OCS and \ce{H2S} molecules trace different type of environments.

As the \ce{H2S}/CS ratio, the SO/CS ratio increases from dense to diffuse ISM. But, only a lower limit on SO is provided toward IRAS 16293-2422 \citep{drozdovskaya2018}. \cite{semenov2018} provided upper limit on the SO column density in disks (i.e. toward DM Tau), giving an SO/CS ratio lower than the lower limit derived toward dense core \citep{vastel2018}. Contrary to the SO/CS, the \ce{SO2}/CS ratio appears rather flat across the different environments compared here, and so do not seem to characterize any specific S-chemistry behavior that would depend on its physical environment. PDR value is found compatible with cold core trend, and with the IRAS 16293-2422 value but to a lesser extent regarding the large uncertainties found. Similarly, the upper limit set by \cite{semenov2018} do not allow us to settle any specific trend for this ratio. 

Finally, the \ce{C2S}/CS ratio decreases from dense to diffuse ISM, in an even more pronounced way than the OCS/CS and \ce{H2S}/CS ratios do. In this case, the disk upper limit \citep{chapillon2012} found in between the value from the dense prestellar core L1544 \citep{vastel2018} and the Horsehead nebula cold core position (Riviere-Marichalar et al. submitted), thus seems to be characteristic of molecular cloud environments.


On average, the S-bearing molecules observed in PPD would behave more like in cold core or molecular clouds. In particular, Fig~\ref{fig:histo} highlights that CS is one of the most abundant S-species detected so far in the different astrophysical environments discussed in this section. However, additional S-bearing PPD observations are required to test these preliminary conclusions.

\label{subsec:iso-ratio-discussion}

\section{Conclusion}
\label{sec:conclusion}

We presented here new ALMA observations of sulfur-bearing molecules toward a sample of six protoplanetary disks.
Our main findings are summarized below:
\begin{enumerate}
    \item The CS $5-4$ rotational transition was observed and detected toward five Taurus PPDs (DM Tau, DO Tau, CI Tau, LkCa 15, MWC 480), and the CS $6-5$ transition toward three PPDs (LkCa 15, MWC 480 and V4046 Sgr). 
    \item We used the two main CS isotopologue rotational transitions ($6-5$ and $5-4$), toward the MWC~480 and LkCa~15 disks, to derive their respective column densities and rotational temperatures of {\ntot(CS)$\simeq (7.3\pm0.8)\tdix{12}\cc$, $\tex\simeq23\pm2.5$~K} and {\ntot(CS)=$(1.2\pm0.1)\tdix{13}\cc$, $\tex=27\pm2.3$~K}, assuming LTE and optically thin emission.
    \item We used the average CS rotational temperature derived toward MWC~480 and LkCa~15, i.e. $\tex\simeq25\pm5$~K, to estimate the CS column densities toward our full disk sample, resulting in an average CS column density of \ntot(CS)$\approx7\tdix{12}$, with a standard deviation of $\approx2\tdix{11}$.
    \item We report the first detections in disks of $^{13}$CS, C$^{34}$S (in LkCa 15), and \ce{H2CS} (in MWC~480). 
    \item We used the CS isotopologue detections toward LkCa~15 to assess the CS disk opacity and tentatively conclude that either both $^{13}$C and $^{34}$S are enhanced by factors of a few (the former showing stronger enhancement than the latter) or the main CS isotopologue lines are optically thick, or there is substantial sub-structure that causes beam dilution.
    \item From the three \ce{H2CS} line detections obtained toward MWC~480 we derived an \ce{H2CS} column density and rotational temperature of {\ntot(\ce{H2CS})$\simeq 3 \tdix{12}$, $\tex\simeq41$~K}. Toward LkCa~15 we use the tentative line detections to derive a less than a factor of 2 lower column density upper limit.
    \item Comparing our CS disk observations to 2D disk astrochemical modeling shows that the CS chemistry seems to be relatively well understood, with column density predictions within a factor of a few compared to the observations found in both MWC~480 and LkCa~15. 
    \item We also compared our \ce{H2CS} observations to 2D disk astrochemical modeling. For that case, our model underpredict \ce{H2CS} in MWC~480 by 1-2 orders of magnitude  and also incorrectly predicts that it should be more abundant in LkCa~15 than in MWC~480, emphasizing the need for further S-disk theoretical investigations.
    \item Concerning S-carriers in disks, our study agrees with previous work finding that CS constitutes one of the main abundant S-species in disk and that it can be produced, at least partly, in protoplanetary disk stage itself.
    \item Furthermore, the main disk strata - the disk atmosphere, intermediate molecular layers and the disk midplane - are analogs to ISM astrophysical environments including PDRs, protostar molecular envelops and dense core, respectively. When comparing the S-bearing column density ratios observed so far in disks (i.e. from this study and literature ones) with their corresponding values observed in the ISM astrophysical environments aforementioned, we find that S-disk chemistry appears more closely related to the one observed in molecular cloud environments. This is consistent with model results which places the origins of the observed emission in the disk molecular layer.
\end{enumerate}

Going forward, additional S-bearing PPD observations are required to increase the variety of S-species observed in disks and construct robust comparisons between S-disk chemistry and earlier ISM evolutionary stages. Such comparisons are needed both to further characterize what kind of chemistry drives the sulfur chemical evolution in disks, and to assess the degree of inheritance from the natal molecular cloud. 

Higher resolution and better signal-to-noise ratio observations would also help to solve any beam dilution issue and to disentangle whether or not substantial carbon and/or sulfur isotopic enhancements occur at protoplanetary disk stage. Such studies combined with isotopic measurements of sulfur carriers in comets would further aid in addressing the inheritance question.

Additional excitation lines, observed with higher signal-to-noise ratio would allow for non-LTE modelling and therefore better constraints on both the CS and \ce{H2CS} column densities and line emission regions.
Non-LTE modeling would also be helpful to better constrain the CS and \ce{H2CS} observations. For instance, it could help in better identifying in which disk layers these respective S-species dominate. On the astrochemical modeling side, there is a need both for more refined disk structures, and for more complete sulfur chemical networks.

Finally, we are ultimately aiming to address how much of the volatile sulfur will be incorporated into the nascent planets? In addition to better constraints on disk sulfur chemistry, addressing this question requires models that couple chemistry and dynamics from disk formation to planet assembly.
This long-term goal is needed to address the sulfur budgets on planets and thus the planet hospitality to the origins of life \citep{ranjan2018}.

\acknowledgments 
\paragraph{{\it Acknowledgments}}
This paper uses the following ALMA data:  ADS$/$JAO.ALMA\#2016.1.01046.S,\\ ADS$/$JAO.ALMA\#2016.1.00627.S, and\\ ADS$/$JAO.ALMA\#2013.1.01070.S. ALMA is a
partnership of ESO (representing its member  states), NSF (USA), NINS (Japan), together  with  NRC (Canada), NSC, ASIAA (Taiwan) and  KASI (Republic of Korea), in cooperation with the Republic of  Chile. The Joint ALMA Observatory (JAO) is operated by ESO, AUI/NRAO and NAOJ. The  authors  would  like  to  thank  the two  anonymous  referees for  valuable  suggestions  and  comments. RLG thanks Valentine Wakelam for allowing her to make use of the \texttt{Nautilus} code in independent research. This work was supported by an award from the Simons Foundation (SCOL \# 321183, KO). J.B.B. acknowledges funding from the National Science Foundation Graduate Research Fellowship under Grant DGE1144152.

\bibliography{disk-sulfur-paper}

\appendix

\section{Building rotational diagrams from disk-integrated flux densities}
\label{app:Nu_SnuDv}

Assuming optically thin transitions, the disk-integrated flux densities $S_\nu \Delta v$, can be related to the column density of their respective upper energy state, $N_u$, as follows:
\begin{equation}
    N_u= \frac{4\pi S_\nu \Delta v}{A_{ul} \Omega hc},
\label{eq:Nu}
\end{equation}
where $S_\nu$ is the flux density, $\Delta v$ the line width, $A_{ul}$ the Einstein coefficient, $c$ the speed of light and
$\Omega$ the solid angle subtended by the source \citep[e.g.][]{bisschop2008,loomis2018}. For this analysis we use the disk flux densities $S_\nu \Delta v$ integrated out to 150~au and 200~au in the MWC~480 and LkCa~15 disks respectively, referred to R$_\sigma$ in Table~\ref{tab:obs-list}.

The total column density, $N_{\rm{tot}}$, and rotational temperature, $T$, can then be derived from the upper level population, $N_u$, following the Boltzmann distribution:
\begin{equation}
   \frac{N_u}{g_u}= \frac{N_{\rm{tot}}}{Q_{\rm{rot}}(T)}e^{-E_u/k_BT},
    \label{eq:boltzmann_dist}
\end{equation}
with $g_u$ and $E_u$ the degeneracy and energy of the upper energy level $u$, respectively; $k_{\rm{B}}$ the Boltzmann constant; and $Q_{\rm{rot}}$ the partition function of the molecule, which for a diatomic molecule as CS can be approximated by:
\begin{equation}
    Q_{\rm{rot}}(T) \approx \frac{k_B\,T}{h\,B_0}+\frac{1}{3},
    \label{eq:part_func}
\end{equation}
\citep{gordy_cook1984}. $h$ is the Planck constant and $B_0$ the rotational constant for CS.

In practice, this can be done by applying a linear least square regression to the logarithm of Eq.~\ref{eq:boltzmann_dist}, as depicted in blue in Fig.~\ref{fig:CS-rot-dia}, but this fit does not take into account that the lines may be slightly optically thick. This can be included in the fit by multiplying $N_u$ by the "optical depth correction factor" for a square line profile, $C_\tau =\frac{\tau}{1-e^{-\tau}}$, in case $\tau~\cancel{\ll}~1$ \citep{goldsmith1999}.
The resulting logarithm of Eq.~\ref{eq:boltzmann_dist} is then: 
\begin{equation}
   \ln{\frac{N_u}{g_u}}= \ln{N_{\rm{tot}}}-\ln{Q_{\rm{rot}}(T)}-\ln{C_\tau}-\frac{E_u}{k_BT}.
    \label{eq:Nu_gu_corr}
\end{equation}

In the LTE approximation, the optical depth of a given transition at \tex\ can be expressed as a function of its upper energy level population using Eq.~\ref{eq:boltzmann_dist} as follows:
\begin{equation}
   \int \tau_\nu {\rm d}v = \frac{ N_u A_{ul} \, c^3}{8\pi \nu^3}(e^{h\nu/k_B\tex} -1).
   \label{eq:opacity-Nu}
\end{equation}

As mentioned in \S~\ref{subsubsec:pop-dia}, the observed CS lines are optically thin based on their brightness temperatures, i.e $\tau<1$, in which case the line profiles remain Gaussian, and the integral of the opacity is given by:
\begin{equation}
    \int \tau_\nu {\rm d}v = 
    \tau_\nu \Delta v_{\rm{FWHM}} \sqrt{\frac{\pi}{4\ln2}},
    \label{eq:tau_nu_fwhm}
\end{equation}
where $\Delta v_{\rm{FWHM}}=\sqrt{8\ln 2}\,\sigma_v$ is the full width at half maximum of the observed transition. Thus, Eq.~\ref{eq:opacity-Nu} can be rewritten as follows:
\begin{equation}
   \tau_\nu = \sqrt{\frac{4\ln2}{\pi}} \frac{ N_u A_{ul} \, c^3}{\Delta v_{\rm{FWHM}} \, 8\pi \nu^3}(e^{h\nu/k_B\tex} -1).
   \label{eq:int_tau_nu_gauss_prof}
\end{equation}

For optically thin lines, i.e $\tau<1$, the line profiles remain Gaussian, and the integral of their opacity is given by:
\begin{equation}
    \int \tau_\nu {\rm d}v = 
    \tau_\nu \Delta v_{\rm{FWHM}} \sqrt{\frac{\pi}{4\ln2}},
    \label{eq:tau_nu_fwhm}
\end{equation}
where $\Delta v_{\rm{FWHM}}=\sqrt{8\ln 2}\,\sigma_v$ is the full width at half maximum of the observed transition. Thus, Eq.~\ref{eq:opacity-Nu} can be rewritten as follows:
\begin{equation}
   \tau_\nu = \sqrt{\frac{4\ln2}{\pi}} \frac{ N_u A_{ul} \, c^3}{\Delta v_{\rm{FWHM}} \, 8\pi \nu^3}(e^{h\nu/k_B\tex} -1).
   \label{eq:int_tau_nu_gauss_prof}
\end{equation}

Subtracting Eq.~\ref{eq:int_tau_nu_gauss_prof} in $C_\tau$ allows 
the construction of a likelihood function $\mathcal{L}(N_u,\tex)$, that can be used for least square minimization. 

\section{Radiative transfer equation used}
\label{app:rad-tr-eqn}
Assuming  a  uniform  excitation  temperature, \tex, along the line of sight, the spectrum intensity at a given frequency $\nu$, corresponding to the observable source brightness temperature $T_B$, is given by the radiative transfer equation as follows:

\begin{equation}
T_{B} \, = \, f \, [J_{\nu}(T_{ex})-J_{\nu}(T_{bg})] \, [1-e^{-\tau_{\nu}}],
\label{eq:transfert}
\end{equation}
where $f$ represents the filling factor, $T_{bg}$ the background temperature, $\tau$ the optical depth and
\begin{equation}
J_{\nu}(T)=\frac{h\nu/k_{\rm{B}}}{e^{h\nu/k_{\rm{B}}T}-1},
\label{eq:RJ_ture} 
\end{equation}
the Rayleigh-Jeans equivalent temperature, with $k_{\rm{B}}$ and $h$ the Boltzmann and Planck constants, respectively.

\section{Channel maps}
In this section we present the channel maps of all the lines presented in the paper.
 
\begin{figure*}
\centering
\includegraphics[scale=0.64]{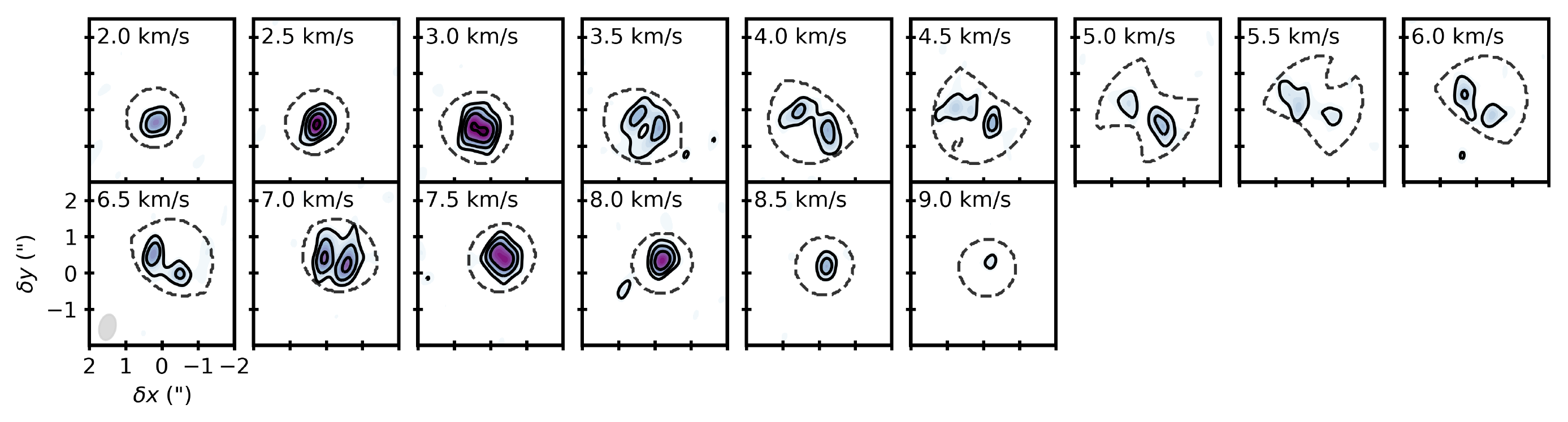}
{\caption{Channel maps of CS $5-4$ line toward MWC~480 with the corresponding Keplerian mask superimposed in dashed black line. Only the emission above $2\times$ the median RMS are shown. The black solid line contours correspond to [3, 6, 10, 15]$\times$ the median RMS.}}
\label{fig:channel-maps-mwc480-cs54}
\end{figure*}

\begin{figure*}
\centering
\includegraphics[scale=0.63]{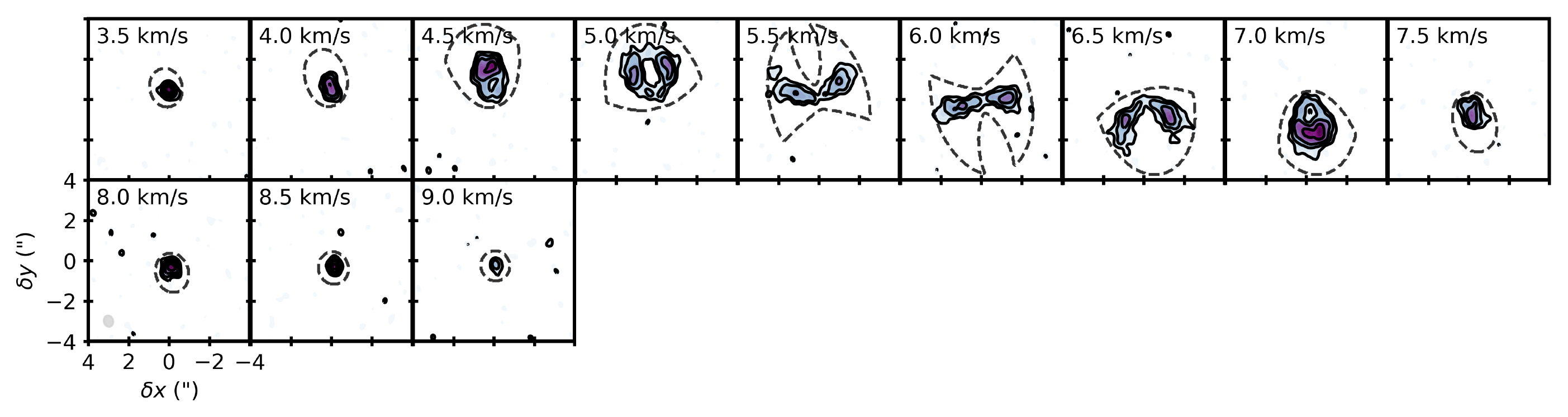}
{\caption{Same as Fig.~\ref{fig:channel-maps-mwc480-cs54} but for CS $5-4$ line toward CI~Tau.}}
\label{fig:channel-maps-citau-cs54}
\end{figure*}

\begin{figure*}
\centering
\includegraphics[scale=0.72]{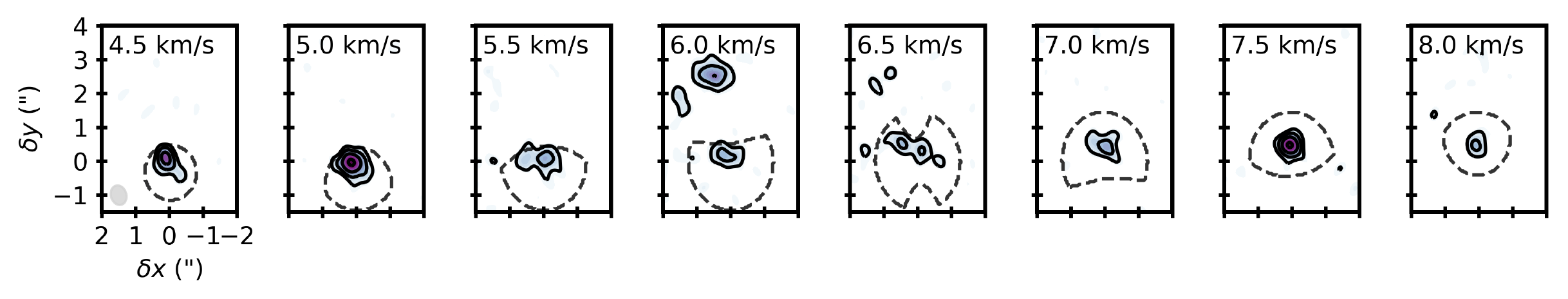}
{\caption{Same as Fig.~\ref{fig:channel-maps-mwc480-cs54} but for CS $5-4$ line toward DO~Tau.}}
\label{fig:channel-maps-dotau-cs54}
\end{figure*}

\begin{figure*}
\centering
\includegraphics[scale=0.64]{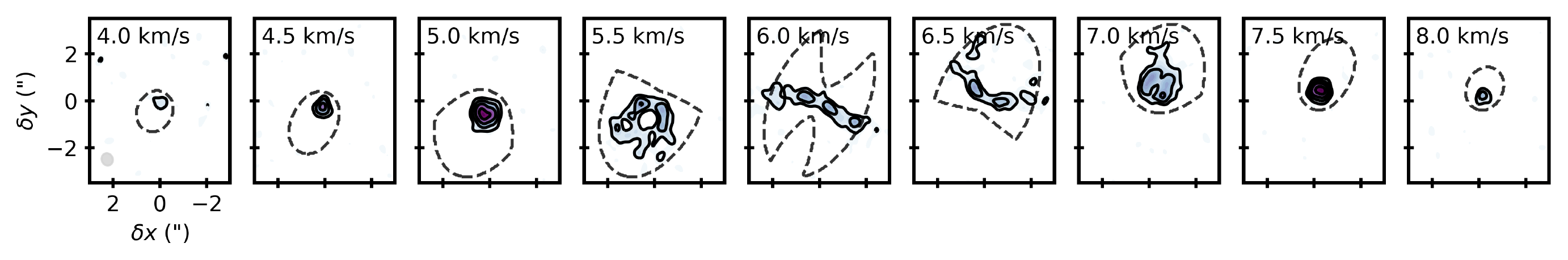}
{\caption{Same as Fig.~\ref{fig:channel-maps-mwc480-cs54} but for CS $5-4$ line toward DM~Tau.}}
\label{fig:channel-maps-dmtau-cs54}
\end{figure*}

\begin{figure*}
\centering
\includegraphics[scale=0.6]{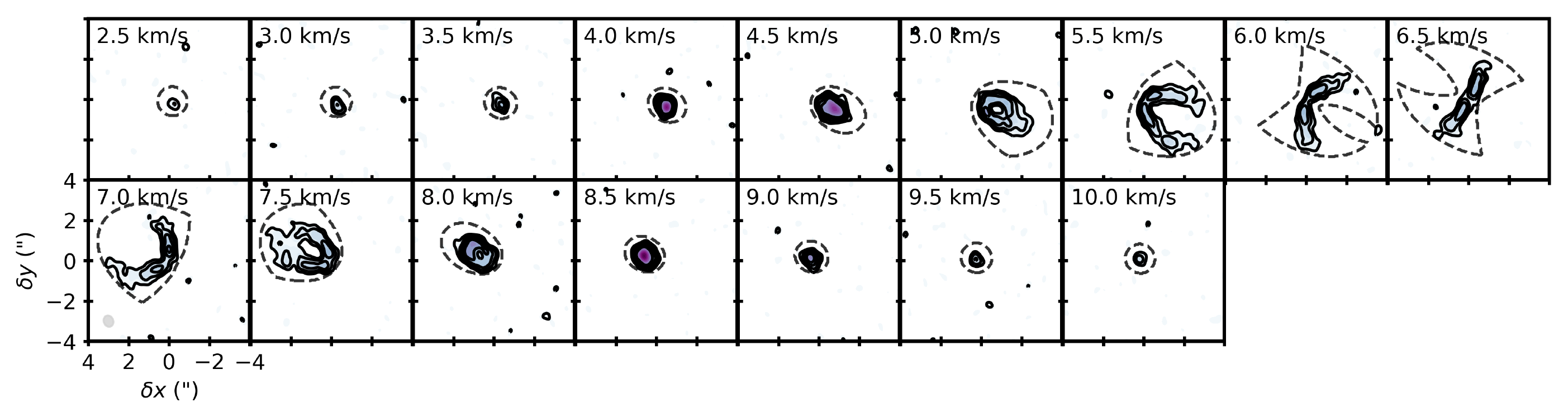}
{\caption{Same as Fig.~\ref{fig:channel-maps-mwc480-cs54} but for CS $5-4$ line toward LkCa~15.}}
\label{fig:channel-maps-LkCa15-cs54}
\end{figure*}

\begin{figure*}
\centering
\includegraphics[scale=0.6]{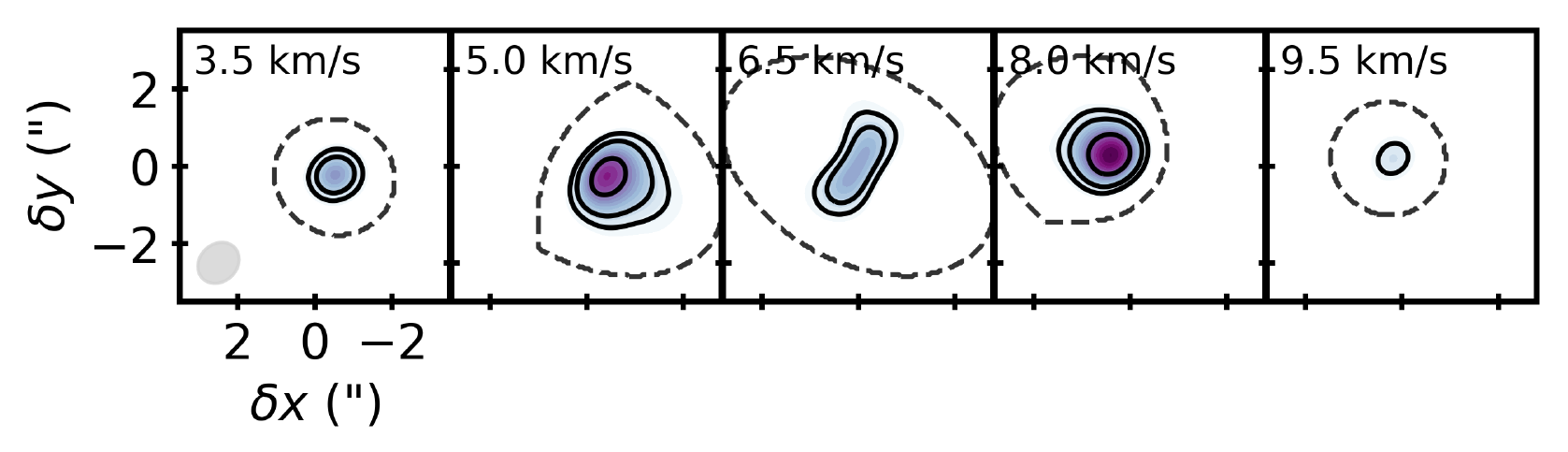}
{\caption{Same as Fig.~\ref{fig:channel-maps-mwc480-cs54} but for CS $6-5$ line toward LkCa~15.}}
\label{fig:channel-maps-LkCa15-cs65}
\end{figure*}

\begin{figure*}
\centering
\includegraphics[scale=0.6]{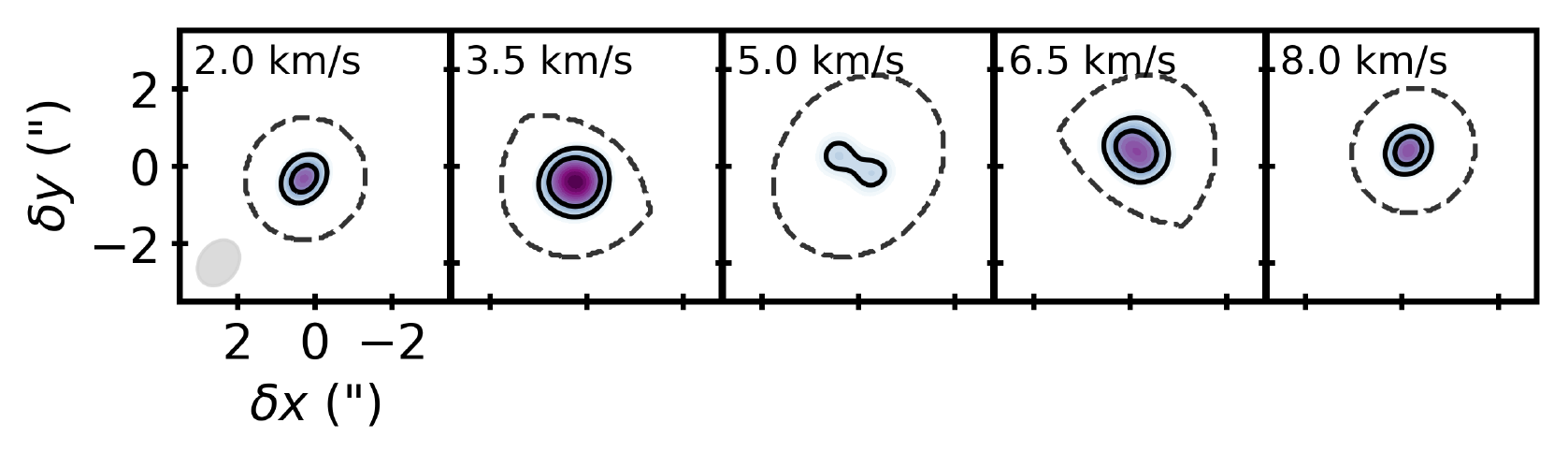}
{\caption{Same as Fig.~\ref{fig:channel-maps-mwc480-cs54} but for CS $6-5$ line toward MWC~480.}}
\label{fig:channel-maps-mwc480-cs65}
\end{figure*}

\begin{figure*}
\centering
\includegraphics[scale=0.6]{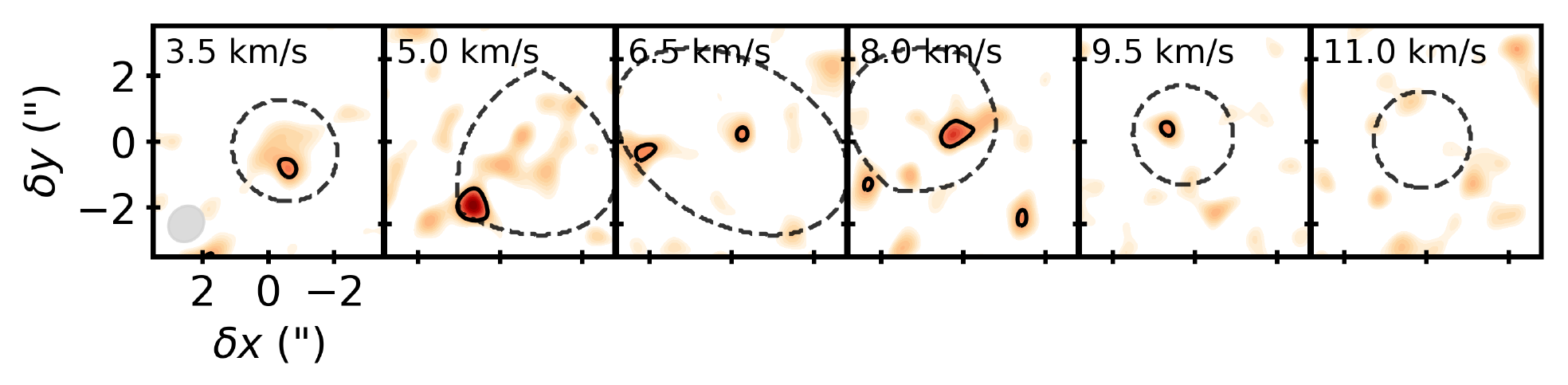}
{\caption{Same as Fig.~\ref{fig:channel-maps-LkCa15-cs65} but for $^{13}$CS $6-5$ line toward LkCa~15 with only the emission above $0.2\times$ the median RMS shown.  The black solid line contour correspond to $0.5\times$ the median RMS.}}
\label{fig:channel-maps-LkCa15-13cs65}
\end{figure*}

\begin{figure*}
\centering
\includegraphics[scale=0.6]{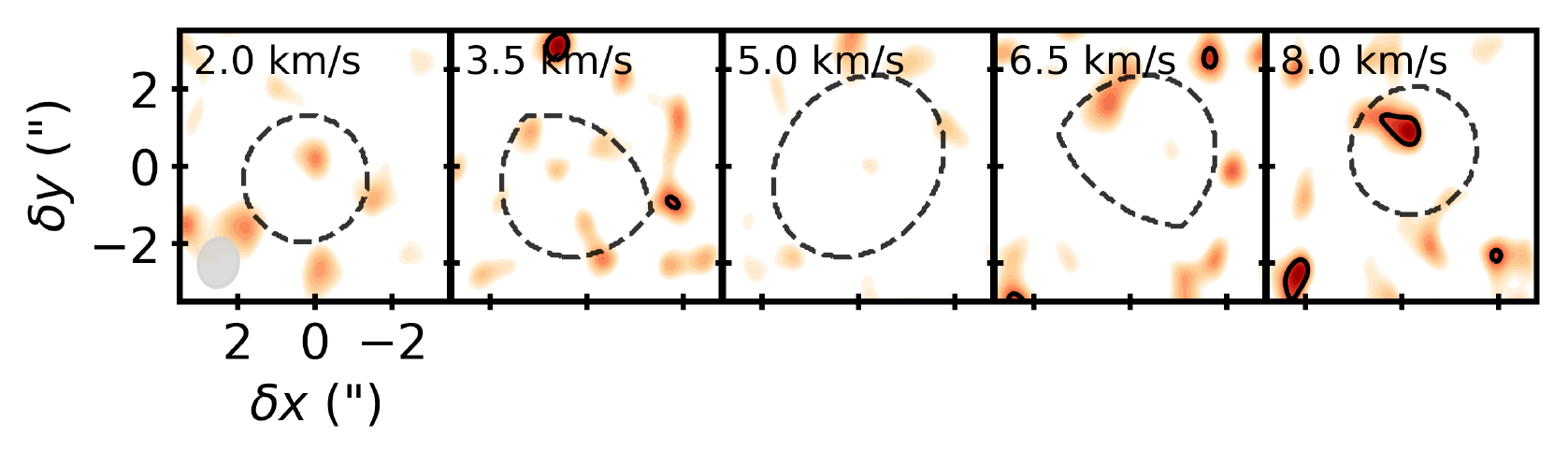}
{\caption{Same as Fig.~\ref{fig:channel-maps-LkCa15-cs65} but for $^{13}$CS $6-5$ line toward MWC~480.}}
\label{fig:channel-maps-mwc480-13cs65}
\end{figure*}

\begin{figure*}
\centering
\includegraphics[scale=0.6]{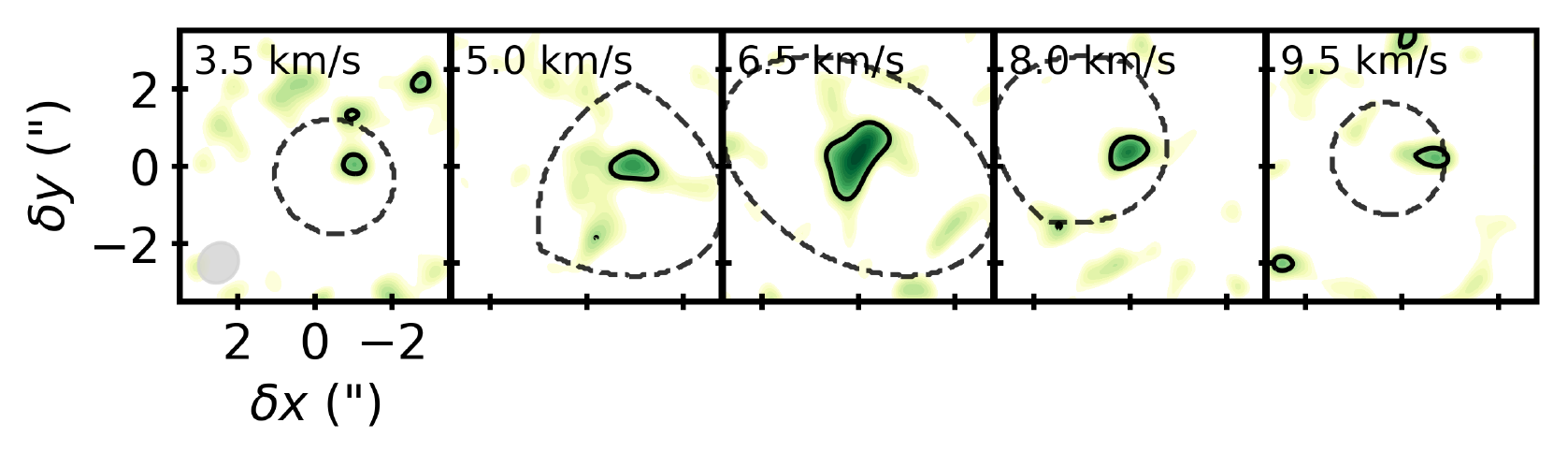}
{\caption{Same as Fig.~\ref{fig:channel-maps-LkCa15-cs65} but for C$^{34}$S $6-5$ line toward LkCa~15.}}
\label{fig:channel-maps-LkCa15-c34s65}
\end{figure*}

\begin{figure*}
\centering
\includegraphics[scale=0.6]{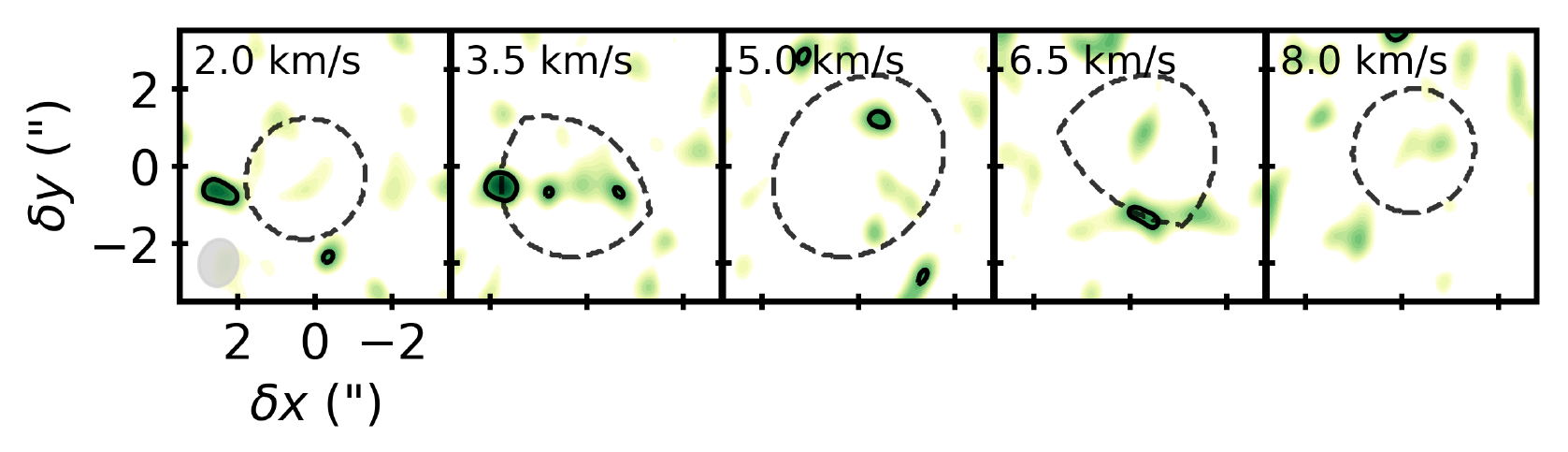}
{\caption{Same as Fig.~\ref{fig:channel-maps-LkCa15-cs65} but for C$^{34}$S $6-5$ line toward MWC~480.}}
\label{fig:channel-maps-mwc480-c34s65}
\end{figure*}

\begin{figure*}
\centering
\includegraphics[scale=0.6]{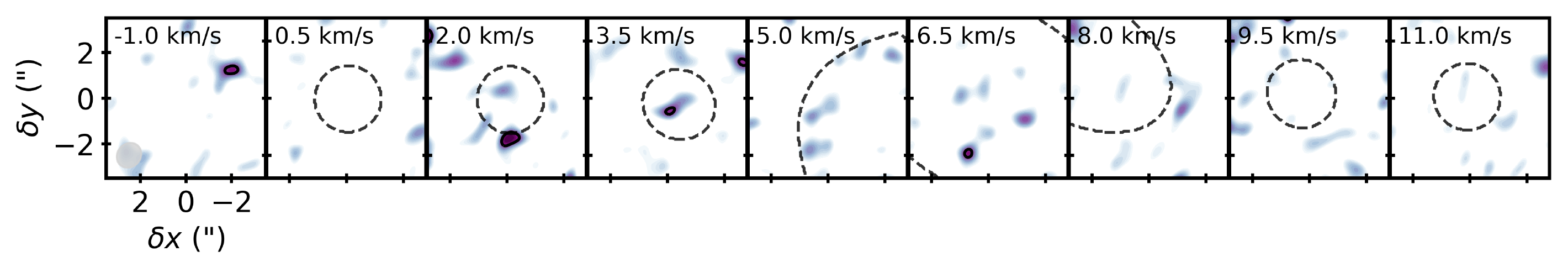}
{\caption{Same as Fig.~\ref{fig:channel-maps-LkCa15-cs65} but for \ce{H2CS} $8_{17}-7_{16}$ line toward LkCa~15.}}
\label{fig:channel-maps-LkCa15-h2cs_817_716}
\end{figure*}

\begin{figure*}
\centering
\includegraphics[scale=0.6]{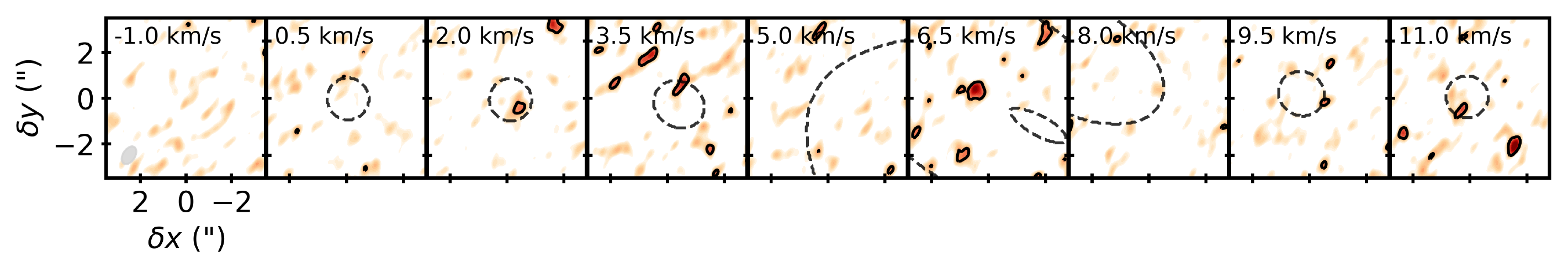}
{\caption{Same as Fig.~\ref{fig:channel-maps-LkCa15-cs65} but for \ce{H2CS} $9_{18}-8_{17}$ line toward LkCa~15.}}
\label{fig:channel-maps-LkCa15-h2cs_918_817}
\end{figure*}

\begin{figure*}
\centering
\includegraphics[scale=0.6]{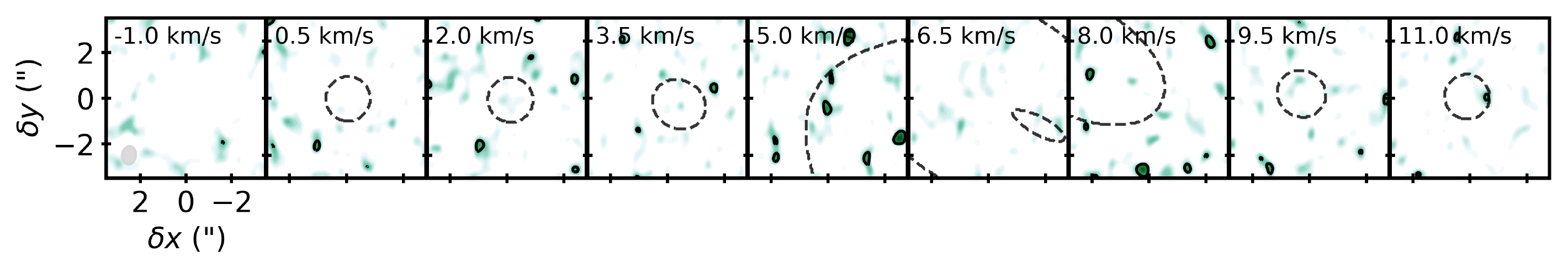}
{\caption{Same as Fig.~\ref{fig:channel-maps-LkCa15-cs65} but for \ce{H2CS} $9_{19}-8_{18}$ line toward LkCa~15.}}
\label{fig:channel-maps-LkCa15-h2cs_919_818}
\end{figure*}

\begin{figure*}
\centering
\includegraphics[scale=0.6]{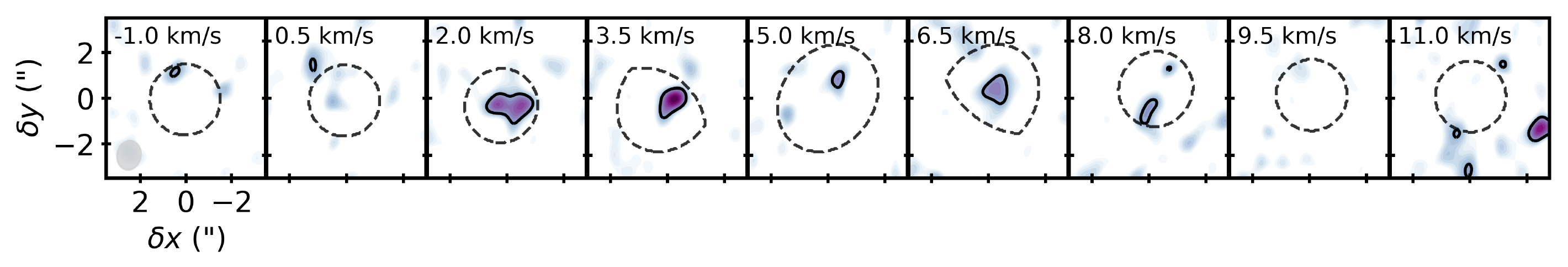}
{\caption{Same as Fig.~\ref{fig:channel-maps-LkCa15-cs65} but for \ce{H2CS} $8_{17}-7_{16}$ line toward MWC~480.}}
\label{fig:channel-maps-mwc480-h2cs_817_716}
\end{figure*}

\begin{figure*}
\centering
\includegraphics[scale=0.6]{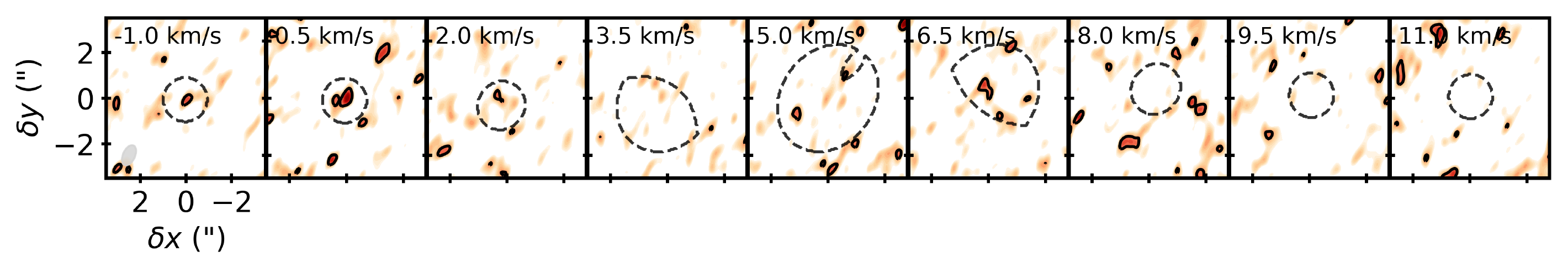}
{\caption{Same as Fig.~\ref{fig:channel-maps-LkCa15-cs65} but for \ce{H2CS} $9_{18}-8_{17}$ line toward MWC~480.}}
\label{fig:channel-maps-mwc480-h2cs_918_817}
\end{figure*}

\begin{figure*}
\centering
\includegraphics[scale=0.6]{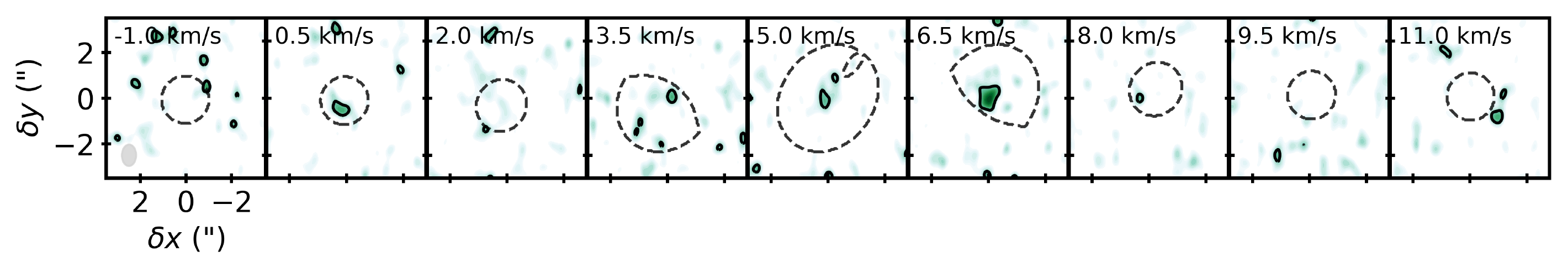}
{\caption{Same as Fig.~\ref{fig:channel-maps-LkCa15-cs65} but for \ce{H2CS} $9_{19}-8_{18}$ line toward MWC~480.}}
\label{fig:channel-maps-mwc480-h2cs_919_818}
\end{figure*}

\section{Modeled fractional abundances}
\label{app:modelled-fractional-ab}

\begin{figure*}
\begin{center}
\includegraphics[scale=0.38]{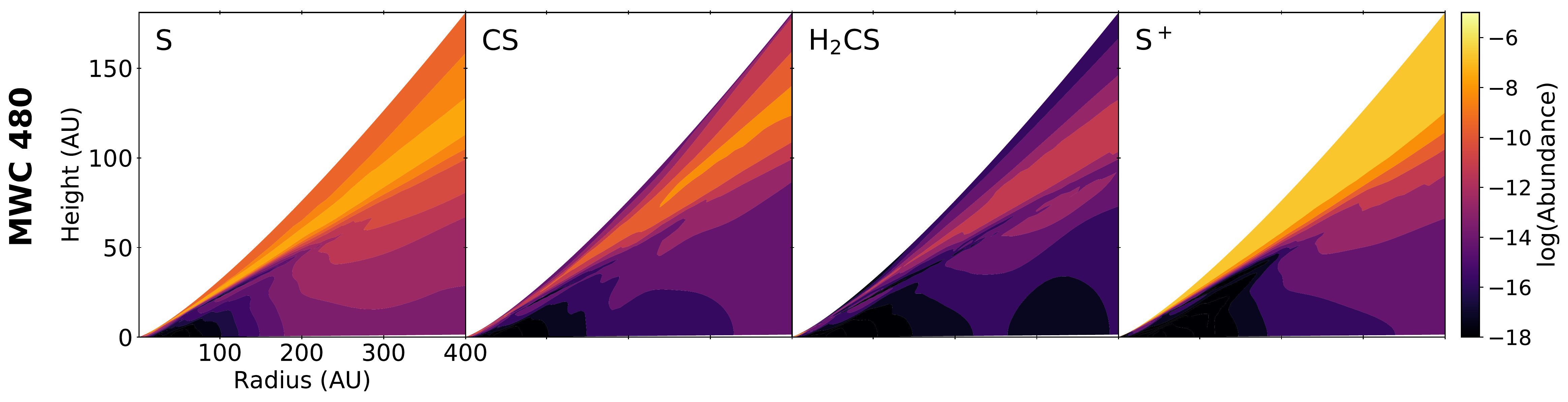}
\includegraphics[scale=0.38]{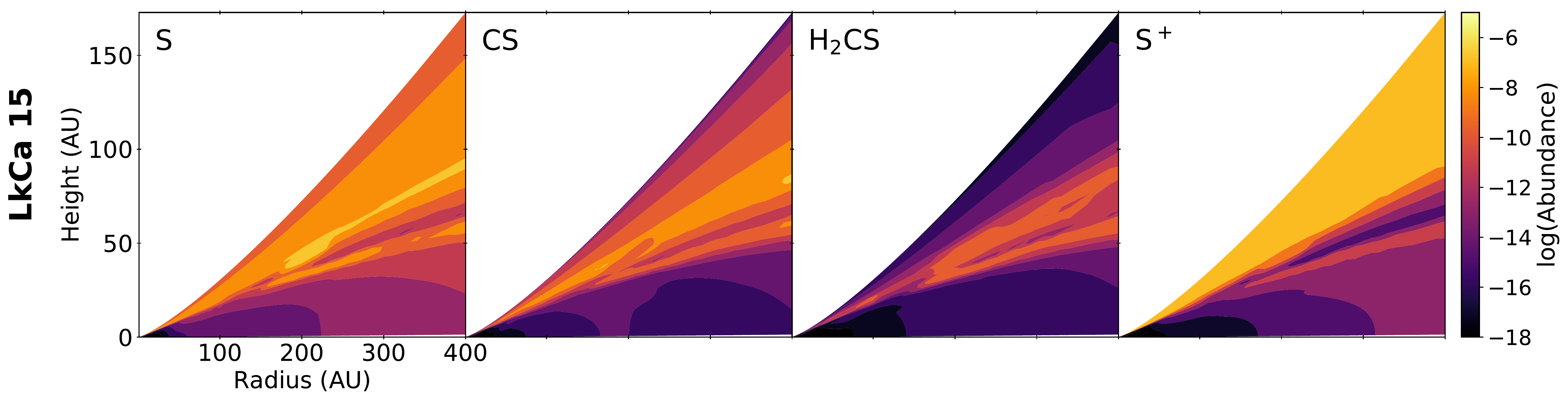}
\caption{Height versus radius {\bf fractional abundances (i.e. the abundances relative to the total abundance of H nuclei)} of CS, \ce{H2CS}, \ce{S+}, and S in our disk models. {\it Top panels:} MWC~480 model, {\it Bottom panels:} LkCa~15 model. \label{fig:2D-modelled-ab}}
\end{center}
\end{figure*}


\software  {\texttt{CASA}} \citep[v4.7.2,][]{mcmullin2007}, {\texttt{Astropy} \citep{astropy}, \texttt{Matplotlib} \citep{matplotlib}, \texttt{NumPy} \citep{numpy_article},  \texttt{emcee} \citep{emcee2013},  \texttt {SciPy} \citep{scipy}, \texttt{scikit-image} \citep{scikit-image}, \texttt{Nautilus-v1.1} \citep{hersant2009,ruaud2016,wakelam2016}}



\end{document}